\documentclass[journal]{IEEEtran}
\usepackage[cmex10]{amsmath}
\usepackage{eqparbox}
\usepackage[usenames, dvipsnames]{color}
\usepackage{mathtools,amssymb,bm,mathabx}
\usepackage{amstext}
\usepackage{amssymb}
\usepackage{graphicx}
\usepackage{color}
\usepackage{booktabs}
\usepackage{longtable}
\usepackage{multicol}
\usepackage{multirow}
\usepackage{amsfonts}
\usepackage{dsfont}
\usepackage{array}
\usepackage[ruled]{algorithm}
\usepackage{algorithmic}
\usepackage{cases}
\usepackage{pgfplots}
\usepackage{accents}
\usepackage{caption}
\usepackage[footnotesize]{subfigure}
\usepackage{amsthm,xparse}
\DeclareCaptionLabelSeparator{periodspace}{.\quad}
\usepackage{float}
\usepackage{cite}
\usepackage [autostyle, english = american]{csquotes}
\usepackage{hyperref}
\theoremstyle{remark}

\newcommand\ASTART{\bigskip\noindent\begin{minipage}[b]{0.5\linewidth}}
	
	\newcommand\AENDSKIP{\end{minipage}\bigskip}
\newcommand\AEND{\end{minipage}}
\ifCLASSOPTIONcaptionsoff
\usepackage[nomarkers]{endfloat}
\let\MYoriglatexcaption\caption
\renewcommand{\caption}[2][\relax]{\MYoriglatexcaption[#2]{#2}}
\fi
\theoremstyle{plain}
\newtheorem{thm}{\textbf{Theorem}}
\newtheorem{lem}{\textbf{Lemma}}

\theoremstyle{definition}
\newtheorem{defn}{\textbf{Definition}}

\theoremstyle{remark}
\newtheorem{rem}{\bf Remark}

\newcommand*{\rom}[1]{\expandafter\@slowromancap\romannumeral #1@}

\newcommand{\RN}[1]{%
\textup{\uppercase\expandafter{\romannumeral#1}}%
}

\usepackage{standalone}
\graphicspath{{./Figures/}} 
\begin{document}
%
%\onecolumn
\title{Multi-weight Matrix Completion with Arbitrary Subspace Prior Information}
\author{Hamideh.Sadat~Fazael~Ardakani, Niloufar~Rahmani, Sajad~Daei}%
%	\thanks{H S. Fazael Ardakani and S. Daei and F. Haddadi are with the School of Electrical Engineering, Iran University of Science \& Technology.}}
\maketitle

\begin{abstract}
	 Matrix completion refers to completing a low-rank matrix from a few observed elements of its entries and has been known as one of the significant and widely-used problems in recent years. The required number of observations for exact completion is directly proportional to rank and the coherency parameter of the matrix.  In many applications, there might exist additional information about the low-rank matrix of interest. For example, in collaborative filtering, Netflix and dynamic channel estimation in communications, extra subspace information is available. More precisely in these applications, there are prior subspaces forming multiple angles with the ground-truth subspaces. In this paper, we propose a novel strategy to incorporate this information into the completion task. To this end, we designed a multi-weight nuclear norm minimization where the weights are such chosen to penalize each angle within the matrix subspace independently. We propose a new scheme for optimally choosing the weights. Specifically, we first calculate an upper-bound expression describing the coherency of the interested matrix. Then, we obtain the optimal weights by minimizing this expression. Simulation results certify the advantages of allowing multiple weights in the completion procedure. Explicitly, they indicate that our proposed multi-weight problem needs fewer observations compared to state-of-the-art methods.
\end{abstract}
\section{Introduction}\label{proof.Matrix completion }
Noisy Matrix completion refers to the task of recovering a low-rank matrix $\bm X \in \mathbb R ^{n\times n}$ with rank $r\ll n$ from few noisy observations  $\bm Y = \mathcal{R} _\Omega (\bm X + \bm E) \in \mathbb{R}^{n \times n } $ \cite{candes2009exact,recht2011simpler}. Here, $\bm{E}$ is the noise matrix and the observation operator $\mathcal{R} _\Omega(\bm{Z})$ for a matrix $\bm{Z}$ is defined as
\begin{align}
& \mathcal{R} _\Omega (\bm Z) := \sum_{i,j=1}^{n} \frac{\epsilon_{ij}}{p_{ij}}\langle \bm Z , \bm{e}_i\bm{e}_j \rangle \bm{e}_i\bm{e}_{j}^{T}, \label{sampling_opr}
\end{align}	
where $p_{ij}$ is the probability of observing $(i,j)$-th element of the matrix, and $\epsilon_{ij}$ is a Bernoulli random variable taking values $1$ and zero with probabilities $p_{ij}$ and $1-p_{ij}$, respectively. While there are many matrix solutions satisfying the observation model, it has been proved that the matrix with lowest rank is unique \cite{candes2010matrix}. Hence, to promote the low-rank feature, the following rank minimization is employed:  
\begin{align}
\underset{\bm Z \in \mathbb{R}^{n \times n}}{\rm min}~ {\rm rank}(\bm Z), \quad{\rm s.t.}~\|\bm Y - \mathcal{R} _\Omega (\bm Z)\|_F\le e , \label{min rank(z)}
\end{align}	
where $e$ is an upper-bound for $\|\mathcal{R} _\Omega(\bm{E})\|_F$.
Since \eqref{min rank(z)} is generally NP-hard and intractable, it is common to replace it with the following surrogate optimization:
\begin{align}
\underset{\bm Z \in \mathbb{R}^{n \times n}}{\rm min}~ \|\bm{Z}\|_*, \quad {\rm s.t.}~\|\bm Y - \mathcal{R} _\Omega (\bm Z)\|_F\le e \label{nuclear_prob}
\end{align}	
where $\| \cdot \|_*$ is called the nuclear norm which computes the sum of singular values of a matrix and is considered as a relaxed version of rank function \cite{recht2010guaranteed}. 

In many applications such as quantum state tomography \cite{gross2010quantum}, MRI \cite{haldar2010spatiotemporal,zhao2010low}, collaborative filtering \cite{srebro2010collaborative}, exploration seismology \cite{aravkin2014fast} and Netflix problem \cite{bennett2007netflix}, there is some accessible prior knowledge about the ground-truth subspaces (i.e. the row and column subspaces of the ground-truth matrix $\bm{X}$). For instance, in Netflix problem, prior evaluations of the movies by the referees can provide prior information about the ground-truth subspaces of Netflix matrix. Further, in sensor network localization \cite{so2007theory}, some information about the position of sensors can be exploited as available prior knowledge (c.f. \cite[Section I]{daei2018optimal} and \cite[Section I.A]{eftekhari2018weighted} for more applications).
The aforementioned prior subspace information often appears in the form of column and row $r'$-dimensional subspaces (denoted by $ \widetilde{\bm{\mathcal{U}}}_{r'} $ and $ \widetilde{\bm{\mathcal{V}}}_{r'} $) forming angles with column and row spaces of the ground-truth matrix $\bm X$, respectively. To incorporate the prior information into the recovery procedure, the following tractable problem for low-rank matrix completion is proposed:
\begin{align}
&\underset{\bm Z \in \mathbb{R}^{n \times n}}{\rm min}~ \|\bm Q_{\widetilde{\bm{\mathcal{U}}}_{r'}} \bm{Z} \bm Q_{\widetilde{\bm{\mathcal{V}}}_{r'}} \|_*  \nonumber \\
&{\rm s.t.}~\|\bm Y - \mathcal{R} _\Omega (\bm Z)\|_F \le e \label{prior_nuclear_prob}
\end{align}	
where
\begin{align}
\bm Q_{\widetilde{\bm{\mathcal{U}}}_{r'}}~:=~\widetilde{\bm{U}}_{r'} \bm \Lambda \widetilde{\bm{U}}_{r'}^H + \bm{P}_{\widetilde{\bm{\mathcal{U}}}_{r'}^{\perp}}, \quad
\bm Q_{\widetilde{\bm{\mathcal{V}}}_{r'}}~:=~\widetilde{\bm{V}}_{r'} \bm \Gamma \widetilde{\bm{V}}_{r'}^H + \bm{P}_{\widetilde{\bm{\mathcal{V}}}_{r'}^{\perp}}
\end{align}	
and $\bm \Lambda$, $\bm \Gamma$ are diagonal matrices whose elements are within the interval $[0,1]$, $\widetilde{\bm{U}}_{r'},\widetilde{\bm{V}}_{r'} \in \mathbb R^{n\times r'}$ indicate some bases for the subspaces $\widetilde{\bm{\mathcal{U}}}_{r'}$ and $\widetilde{\bm{\mathcal{V}}}_{r'}$, respectively. Also, the orthogonal projection matrix is defined by $\bm{P}_{\widetilde{\bm{\mathcal{U}}}_{r'}^{\perp}}:=\bm{I}_{n}-\widetilde{\bm{U}}_{r'} \widetilde{\bm{U}}_{r'}^H$ where $\bm{I}_n$ is the identity matrix of size $n$. So according to the definitions, if $\bm \Lambda = \bm \Gamma = \bm{I}_{r'}$, the problem \eqref{prior_nuclear_prob} reduces to the standard nuclear norm minimization \eqref{nuclear_prob}. The values of matrices $\bm \Lambda$ and $\bm \Gamma$ depend on the precision of our available prior knowledge for each direction (i.e. each column of $\widetilde{\bm{U}}_{r'}$) in the form of principal angles. As an example, when the principal angle between the estimated and true basis vectors (or directions) increases, the accuracy of that direction estimate is reduced, and so the assigned weight to that direction estimate should intuitively be large and near $1$.

\subsection{Contributions}
In this paper, we propose a general scheme for low-rank matrix completion with prior subspace information. Since we penalize the inaccuracy of each basis (direction) in the prior subspace in our proposed method, more degrees of freedom are provided in compared to the previous related works and this leads to fewer required observations for matrix completion. Moreover, we design an optimal strategy to promote the prior formation so as to reduce the required number of samples for completion up to the greatest possible extent. This is accomplished by assigning dedicated weights to penalize the bases of the prior subspaces in an optimal manner. Our theoretical and numerical results also certify that our devised method needs fewer samples for matrix completion compared to the existing similar methods in \cite{eftekhari2018weighted} and \cite{ardakani2020multi}.
\subsection{Prior Arts and Key Differences}\label{sec.relatedwork}
In this section, we summarize some of the related existing approaches for completing low-rank matrix. The
authors in \cite{rao2015collaborative} propose a weighted form of trace-norm regularization that outperforms the unweighted version:
\begin{align}\label{5}
\|\bm X \|_{\rm tr} := \|{\rm diag}(\sqrt{\bm p})\bm X {\rm diag}(\sqrt{\bm q}) \|_*
\end{align}
in which $p(i)$ and $q(j)$ indicate the probabilities of the i-th row and the j-th column of the matrix under observation, respectively.

In \cite{angst2011generalized}, \cite{jain2013provable} and \cite{xu2013speedup}, the directions in the row and column subspaces of $\bm{X}$ are penalized based on prior information. In \cite{mohan2010reweighted}, the authors discuss the problem of minimizing re-weighted trace norm as an iterative heuristic and analyze its convergence. In \cite{rao2015collaborative} and \cite{srebro2010collaborative}, the authors considered a generalized nuclear norm to incorporate structural prior information into matrix completion and proposed a scalable algorithm based on the approach of \cite{zhou2012kernelized}.  

Aravkin et al. in \cite{aravkin2014fast} for the first time incorporated prior subspace information into low-rank matrix completion by an iterative algorithm in order to solve:
\begin{align}\label{6}
\underset{\bm Z \in \mathbb{R}^{n\times n}}{\rm min} ~ \| \bm Q_{\widetilde{\bm{\mathcal{U}}}_{r}} \bm Z \bm Q_{\widetilde{\bm{\mathcal{V}}}_{r}}\|_* , \quad {\rm s.t.}~\bm Y = \mathcal{A} (\bm Z) \nonumber
\end{align}
where 
\begin{align}
\bm Q_{\widetilde{\bm{\mathcal{U}}}_{r}}~:=~ \lambda \bm{P}_{\widetilde{\bm{\mathcal{U}}}_{r}^{\perp}} + \bm{P}_{\widetilde{\bm{\mathcal{U}}}_{r}^{\perp}}, \quad \bm Q_{\widetilde{\bm{\mathcal{V}}}_{r}}~:=~ \gamma \bm{P}_{\widetilde{\bm{\mathcal{V}}}_{r}^{\perp}} + \bm{P}_{\widetilde{\bm{\mathcal{V}}}_{r}^{\perp}}
\end{align}	
and $\lambda$, $\gamma$ depend on the maximum principal angle.

Eftekhari et al. in \cite{eftekhari2018weighted} proves that the number of required observations can be reduced compared to the standard nuclear norm minimization in the presence of prior information. Their approach assigns the whole prior subspaces with a single weight that is chosen by maximizing the coherence of the interested matrix. Penalizing the whole subspaces with a single weight seems to be not reasonable since the directions within subspaces have different angles with those of the ground-truth matrix. Thus, it would be better to penalize the far directions more while encouraging the close ones via assigning multiple weights. Further, their approach only works when the prior subspaces are sufficiently close to the ground-truth ones. In \cite{ardakani2020multi}, the authors propose a similar weighting strategy as in \eqref{prior_nuclear_prob} for low-rank matrix recovery problem. Their approach for choosing the weights is to weaken the restricted isometry property (RIP) condition of the measurement operator and is completely different from what the current work will offer. Unfortunately, many measurement operators (including matrix completion framework) fail to satisfy RIP and thus this approach can not be applied to matrix completion problem. Further, unlike \cite{eftekhari2018weighted}, the considered prior subspaces in \cite{ardakani2020multi} can be either far or close to the ground-truth subspaces. In other words, \cite{ardakani2020multi} has shown that far prior subspaces can be beneficial in improving the performance of low-rank matrix recovery. There are also some works of different flavor in matrix completion such as \cite{fathi2021two,ardakani2019greedy} which amounts to designing algorithms for recovering coherent low-rank matrices. Despite these efforts, it is still vague to what extent does prior knowledge help (or hurt) matrix completion.

Another related work with the same model as \eqref{6} is \cite{daei2018optimal} where optimal weights are designed based on statistical dimension theory in full contrast to other works that deal with the RIP bound. However, statistical dimension theory is not applicable to many measurement models such as matrix completion framework. It is worth mentioning that our interest in weighted matrix completion is inspired by a closely related field known as compressed sensing \cite{donoho2006compressed}, \cite{candes2008restricted}, which Needell et al. in \cite{needell2017weighted} present recovery conditions for weighted $\ell_1$-minimization when there are several available prior information about the support of a sparse signal. This prior information is organized into several sets where each contributes to the support with a certain degree of precision and these sets are assigned non-uniform weights. It is equally important to note the term \lq\lq non-uniform weights\rq\rq points out to multiple different weights which are employed interchangeably in this paper. This work can actually be regarded as an extension of \cite{needell2017weighted} to the matrix completion case. In order to penalize different directions of the ground-truth matrix, we use non-uniform weights. Despite the general idea, our used tools and analysis in the current work substantially differs from those in \cite{needell2017weighted} and involves highly challenging and non-trivial mathematical steps. 
\subsection{Outline and Notations}
The paper is organized as follows: In Section \ref{sec.singleweight}, we review the uniform (single) weighted strategy introduced in \cite{eftekhari2018weighted} with more details. In Section \ref{sec.main result}, we present our main results which amounts to proposing a non-uniform weighted nuclear norm minimization. In Sections \ref{sec.simulation}, some numerical results are provided which verify the superior performance of our method. Finally, in Section \ref{sec.conclusion}, the paper is concluded.

Throughout the paper, scalars are indicated by lowercase letters, vectors by lowercase boldface letters, and matrices by uppercase letters. The trace and hermitian of a matrix are shown as ${\rm Tr}(\cdot)$ and $(\cdot)^{\rm H}$, respectively. The Frobenius inner product is defined as $\langle\bm A , \bm B\rangle_F = {\rm Tr}(\bm A \bm{B}^{H})$. $\| \cdot \|$ denote the spectral norm and $\bm X \succcurlyeq 0$ means that $\bm X$ is a semidefinite matrix. We describe the linear operator $\mathcal{A}:\mathbb{R}^{m \times n}\longrightarrow \mathbb{R}^p$ as
$\mathcal{A} \bm X = [\langle \bm X , \bm{A}_1 \rangle_F , \cdots ,\langle \bm X , \bm{A}_p \rangle_F]^T$
where $\bm{A}_i\in \mathbb{R}^{m \times n}$. The adjoint operator of $\mathcal{A}$ is defined as $\mathcal{A}^*\bm y = \sum_{i=1}^{p} y_i\bm{A}_i$ and $\mathcal{I}$ is the identity linear operator i.e. $\mathcal{I}\bm X=\bm X$.

\section{Single Weight Nuclear Norm Minimization}\label{sec.singleweight}
In this section, we explain the strategy of single weight penalization employed in \cite{eftekhari2018weighted}.

\begin{defn}[\cite{daei2018optimal}] Assume $\bm{\mathcal{U}} \in \mathbb{R}^n$ is an $r$-dimensional subspace and $\bm{P}_{\mathcal{U}}$ indicates orthogonal projection onto that subspace, then the coherency of $\bm{\mathcal{U}}$ is defined as:
	\begin{align}\label{coherency}
	\mu(\bm{\mathcal{U}}):=\frac{n}{r}\underset{1\leq i\leq n}{\rm max}~\| \bm{P}_{\mathcal U} \bm{e}_i \|^2,
	\end{align}
	where $\bm{e}_i\in\mathbb{R}^{n}$ is the canonical vector having $1$ in the $i$-th location and zero elsewhere.
	In order to define principal angles between subspaces $\bm{\mathcal U}$ and $\bm{\widetilde{\mathcal U}}$, let $r$ and $r'$ represent dimensions of $\bm{\mathcal U}$ and $\bm{\mathcal{\widetilde U}}$, respectively with $r \leq r'$. There exists $r$ non-increasing principal angles $\theta_u \in [0^o,90^o]^r$
	\begin{align}
	&\theta_u(i) = {\rm min} \bigg\{{\rm cos}^{-1}(\frac{|\langle\bm u , \bm{\widetilde{u}}\rangle|}{\| \bm u \|_2 \| \bm{\widetilde{u}} \|_2}) : \bm u \in \bm{\mathcal{U}} , \bm{\widetilde{u}} \in \bm{\widetilde{\mathcal U}} \nonumber \\
	&\bm u \perp \bm{u}_j ~,~ \bm{\widetilde{u}} \perp \bm{\widetilde{u}}_j ~:~ \forall j \in \{i+1, \cdots , r\} \bigg\}
	\end{align}
	where $\bm u$ and $\bm{\widetilde u}$ indicate principal vectors and $\theta_u(1)$ denotes the maximum principal angle.
\end{defn}
\begin{thm}[\cite{recht2010guaranteed}]
	 Assume $\bm{X}_r=\bm{U}_r\bm{\Sigma}_r\bm{V}_{r}^{H} \in \mathbb R ^{n \times n}$ be a truncated SVD from matrix $\bm X \in \mathbb R ^{n\times n}$, for an integer $r \leq n$ and let $\bm X _{r^+}=\bm X - \bm X_r$ indicates the residual. Consider $\widetilde{\bm{\mathcal U}}_r$ and $\widetilde{\bm{\mathcal V}}_r$ as prior subspace information of $\bm{\mathcal U}_r= {\rm span} (\bm{X}_r)$ and $\bm{\mathcal V}_r={\rm  span}(\bm{X}_r^H)$, respectively. Assume $\eta(\bm X_r)=\eta(\bm U_r \bm V_r^H)={\rm max}_i~\mu_i(\bm{\mathcal U}_r) \vee {\rm max}_j~\nu_j(\bm{\mathcal V}_r) $ indicates the coherence of $\bm X_r$. Additionally, let $\breve{\bm U}$ and $\breve{\bm V}$ represent orthonormal bases for ${\rm span}([\bm{U}_r , \widetilde{\bm{U}}_r])$ and ${\rm span}([\bm V_r , \widetilde{\bm V}_r])$, respectively. For $\lambda,\gamma \in (0,1]$, if $\widehat{\bm{X}}$ is a solution to \eqref{prior_nuclear_prob}, then 
	 \begin{align}
	 &\| \hat{\bm X} - \bm X \|_F \lesssim \frac{\| \bm{X}_{r^+} \|_*}{\sqrt{p}}+e\sqrt{pn} \label{theory_armin} 	 
	 \end{align}	
	 provided that
	 \begin{align}
	 &1 \geq p \gtrsim {\rm max} [{\rm log}(\alpha_1 .n ),1]. \frac{\eta(\bm{X}_r) r {\rm log}n}{n}.\nonumber\\
	 & \quad ~~~~~~~~~~~{\rm max}[1+\frac{\eta(\breve{\bm U} \breve{\bm V}^H)}{\eta(\bm U {\bm V}^H)},1] \nonumber 	 \\
	 &\alpha_3 \leq \frac{1}{8}  \label{condition_armin}
	 \end{align}	
	 where $\eta(\breve{\bm U} \breve{\bm V}^H)$ represents the coherence of $\breve{\bm U} \breve{\bm V}^H$ and
	 \begin{align}
	 &\alpha_1 := \sqrt{ \frac{\lambda^4 {\rm cos}^2 \theta_u(1) + {\rm sin}^2\theta_u(1)}{\lambda^2 {\rm cos}^2 \theta_u(1) + {\rm sin}^2\theta_u(1)} } \sqrt{ \frac{\lambda^4 {\rm cos}^2 \theta_v(1) + {\rm sin}^2\theta_v(1)}{\lambda^2 {\rm cos}^2 \theta_v(1) + {\rm sin}^2\theta_v(1)} } \nonumber\\
	 &\alpha_2 := \bigg( \sqrt{ \frac{\lambda^2 {\rm cos}^2 u + {\rm sin}^2 u}{\gamma^2 {\rm cos}^2 v + {\rm sin}^2 v} } + 
	  \sqrt{ \frac{\gamma^2 {\rm cos}^2 v + {\rm sin}^2 v}{\lambda^2 {\rm cos}^2 u + {\rm sin}^2 u} } \bigg) . \nonumber\\
	&\big( \sqrt{\lambda^4 {\rm cos}^2 u + {\rm sin}^2 u} + \sqrt{\gamma^4 {\rm cos}^2 v + {\rm sin}^2 v} \big) \nonumber \\
	&\alpha_3 := \frac{3\sqrt{1-\lambda^2}~{\rm sin}u}{2\sqrt{\lambda^2 {\rm cos}^2 u + {\rm sin}^2 u}} + 
	\frac{3\sqrt{1-\gamma^2}~{\rm sin}v}{2\sqrt{\gamma^2 {\rm cos}^2 v + {\rm sin}^2 v}}. \nonumber
	 \end{align}
\end{thm}
\begin{rem}
By taking $\lambda = \gamma = 1$, the problem \eqref{prior_nuclear_prob} reduces to the standard unweighted nuclear norm minimization problem \eqref{nuclear_prob}, i.e. it leads to $\alpha_1 = 1$, $\alpha_2 = 4$ and $\alpha_3 = 0$. Thus, \eqref{condition_armin} will change to 
\begin{align}
1 \geq p \gtrsim \frac{\eta(\bm{X}_r)r{\rm log}^2 n }{n}.\big( 1+\sqrt{\frac{\eta(\breve{\bm U} \breve{\bm V}^H)}{\eta(\breve{\bm {U}}_r \breve{\bm V}_r^H)}} \big)
\end{align}
	\end{rem}
Further, due to the term $ \sqrt{\tfrac{\eta(\breve{\bm U} \breve{\bm V}^H)}{\eta(\breve{\bm {U}}_r \breve{\bm V}_r^H)}}$, the probability of an element being observed is worse than \cite{candes2010matrix} for solving \eqref{nuclear_prob} in the noisy case which is considered as $1\geq p \gtrsim \tfrac{\eta (\bm{X}_r)r{\rm log}^2 n}{n}$.
\section{Non-Uniform Weighting}\label{sec.main result}
In this section, we generalize the single (or uniform) weighted matrix completion with nuclear norm minimization approach \cite{eftekhari2018weighted}, to the non-uniform weights. Consider $\bm{\widetilde{\mathcal U}}_{r'}$ and $\bm{\widetilde{\mathcal V}}_{r'}$ as prior subspace information forming angles with $\bm{\mathcal U}_{r}$ and $\bm{\mathcal V}_{r}$, respectively. We optimize the weights according to values of the principal angels. Theorem \eqref{thm2} below provides performance guarantees for both noiseless and noisy matrix completion and uniform weighting strategy is a special case of it. 
\begin{thm} \label{thm2}
	Consider $\bm{X}_r \in \mathbb{R}^{n\times n}$ as a truncated SVD from matrix $\bm X $, and let $\bm X _{r^+}=\bm X - \bm X_r$ indicate the residual. Let ${\bm{\mathcal U}}_r = {\rm span}(\bm{X}_r)$ and ${\bm{\mathcal V}}_r = {\rm span}(\bm{X}_r^H)$ indicate the column and row subspaces of $\bm{X}_r$, and the $r'$-dimensional subspaces  $\bm{\widetilde{\mathcal U}}_{r'}$ and  $\bm{\widetilde{\mathcal V}}_{r'}$ be their corresponding prior subspace estimates, respectively. For each pair of subspaces, consider the non-increasing  principal angle vectors defined bellow to indicate the accuracy of prior information:
	\begin{align}
	\bm{\theta}_u=\angle [\bm{\mathcal{U}},\bm{\widetilde{\mathcal{U}}}],~\bm{\theta}_v=\angle [\bm{\mathcal{V}},\bm{\widetilde{\mathcal{V}}}]
	\end{align}
	Assume $\eta(\bm X_r)=\eta(\bm U_r \bm V_r^H)={\rm max}_i \mu_i~(\bm{\mathcal U}_r) \vee {\rm max}_j~\nu_j(\bm{\mathcal V}_r) $ indicate the coherence of $\bm X_r$. Additionally, let $\breve{\bm U}$ and $\breve{\bm V}$ represent orthonormal bases for ${\rm span}([\bm{U}_r , \widetilde{\bm{U}}_r'])$ and ${\rm span}([\bm V_r , \widetilde{\bm V}_r'])$, respectively. For $\Lambda_{i,i}, \Gamma_{i,i} \in (0,1]$, let $\hat{\bm{X}}$ be a solution to \eqref{prior_nuclear_prob}. Then,
	\begin{align}
	&\| \hat{\bm X} - \bm X \|_F \lesssim \frac{\| \bm{X}_{r^+} \|_*}{\sqrt{p}}+e\sqrt{pn} \label{theory_main} 	 
	\end{align}	
	provided that
	\begin{align}\label{eq.prob_observed}
	&1 \geq p \gtrsim {\rm max}[{\rm log}(\alpha_4.n),1].\frac{\mu(\bm{X}_r)r{\rm log}n}{n}. \nonumber\\
	&	 \quad \quad \quad \quad ~~~~~~~~~~~~~~{\rm max}\big[\alpha_5^2\big(1+\frac{\eta(\breve{\bm U} \breve{\bm V}^H)}{\eta(\bm U {\bm V}^H)}\big),1\big]\nonumber\\
	&\alpha_6\leq\frac{1}{4}
	\end{align}	
	where $\eta(\breve{\bm U} \breve{\bm V}^H)$ is the coherence of $\breve{\bm U} \breve{\bm V}^H$ and
	\begin{align}
	&\alpha_4:=\alpha_4(u_i,v_i,\lambda(i),\gamma_1(i)) := \nonumber \\
	& \sqrt{\underset{i}{\rm max}\bigg(\frac{\lambda_1(i)^4 {\rm cos}^2 \theta_u(i) + {\rm sin}^2\theta_u(i)} {\lambda_1(i)^2 {\rm cos}^2 \theta_u(i) + {\rm sin}^2\theta_u(i)} \bigg)} ~\cdot \nonumber \\
	& \quad \quad \quad \quad \quad  \sqrt{\underset{i}{\rm max}\bigg(\frac{\gamma_1(i)^4 {\rm cos}^2 \theta_v(i) + {\rm sin}^2\theta_v(i)} {\gamma_1(i)^2 {\rm cos}^2 \theta_v(i) + {\rm sin}^2\theta_v(i)}\bigg)}  \nonumber \\
	& \alpha_5 := \alpha_5(\theta_u(i),\theta_v(i),\lambda_1(i),\gamma_1(i)) :=  \nonumber \\
	& \sqrt{\underset{i}{\rm max} (\lambda_1(i)^2{\rm cos}^2\theta_u(i) + {\rm sin}^2 \theta_u(i))} ~\cdot  \nonumber \\
	& \quad \quad \quad \quad \quad\sqrt{\underset{i}{\rm max} \bigg(\frac{\gamma_1(i)^4 {\rm cos}^2 \theta_v(i) + {\rm sin}^2\theta_v(i)} {\gamma_1(i)^2 {\rm cos}^2 \theta_v(i) + {\rm sin}^2\theta_v(i)}\bigg)} + \nonumber \\
	& \sqrt{\underset{i}{\rm max} (\gamma_1(i)^2{\rm cos}^2\theta_v(i) + {\rm sin}^2 \theta_v(i))} ~\cdot  \nonumber \\
	&\quad \quad \quad \quad \quad \sqrt{\underset{i}{\rm max} \bigg(\frac{\lambda_1(i)^4 {\rm cos}^2 \theta_u(i) + {\rm sin}^2\theta_u(i)} {\lambda_1(i)^2 {\rm cos}^2 \theta_u(i) + {\rm sin}^2\theta_u(i)}\bigg)} \nonumber \\ 
	& \alpha_6 := \alpha_6 (\theta_u(i),\theta_v(i),\lambda_1(i),\lambda_2(i),\gamma_1(i),\gamma_2(i)) :=\nonumber \\
	& \sqrt{\underset{i}{\rm max} \bigg( \frac{(1-\lambda_1(i)^2)^2{\rm cos}^2\theta_u(i)+{\rm sin}^2\theta_u(i)}{\lambda_1(i)^2{\rm cos}^2\theta_u(i)+{\rm sin}^2\theta_u(i)} \bigg)} ~\cdot \nonumber \\
	& \quad \quad \quad \quad \quad \sqrt{ \underset{i}{\rm max} \bigg( \frac{(1-\gamma_1(i)^2)^2{\rm cos}^2\theta_v(i)+{\rm sin}^2\theta_v(i)}{\gamma_1(i)^2{\rm cos}^2\theta_v(i)+{\rm sin}^2\theta_v(i)} \bigg) } - \nonumber \\
	& {\rm max}\bigg\{\underset{i}{\rm max}(\lambda_2(i)-1),\underset{i}{\rm max}\bigg( \frac{\lambda_1(i)}{\sqrt{\lambda_1(i)^2{\rm cos}^2\theta_u(i)+{\rm sin}^2\theta_u(i)}}\nonumber\\
	&-1 \bigg)\bigg\} -  {\rm max}\bigg\{\underset{i}{\rm max}(\gamma_2(i)-1), \nonumber \\
	&\quad \quad \quad \quad \quad\underset{i}{\rm max}\bigg( \frac{\gamma_1(i)}{\sqrt{\gamma_1(i)^2{\rm cos}^2\theta_v(i)+{\rm sin}^2\theta_v(i)}}-1 \bigg)\bigg\}.
	\end{align}
\end{thm}
Proof. See Appendix \ref{proof_th}.
\begin{rem}
	If $\bm \Lambda=\bm \Gamma=\bm{I}_{r'}$ then $\bm{Q}_{\bm{\widetilde{\mathcal{U}}_{r'}}} = \bm{Q}_{\bm{\widetilde{\mathcal{V}}_{r'}}} = \bm{I}_{n}$ and the problem reduces to the standard matrix completion \eqref{prior_nuclear_prob}. Also, considering $\bm \Gamma=\gamma \bm{I}_r$ and $\bm \Lambda=\lambda \bm{I}_r$, \eqref{prior_nuclear_prob} reduces to the single weighted problem studied in \cite{eftekhari2018weighted}.
\end{rem}
\begin{rem}(Finding optimal weights)
	Our goal is to reduce the number of required observations or alternatively the recovery error in matrix completion problem. Therefore, we choose the optimal weights that minimizes the lower-bound on $p$ in \eqref{eq.prob_observed} and consequently the recovery error in \eqref{theory_main}.
\end{rem}
\begin{rem}
	 In our proposed model, all principal angels between subspaces are assumed to be accessible. Accordingly, this provides higher degrees of freedom to reduce the number of required observations which is in turn translated to enhance the performance accuracy.
\end{rem}

\section{Simulation Results}\label{sec.simulation}
\begin{figure*}
	%	\hspace*{-.5cm}
%	\centering
	\includegraphics[width=3.5in,height=1.5in]{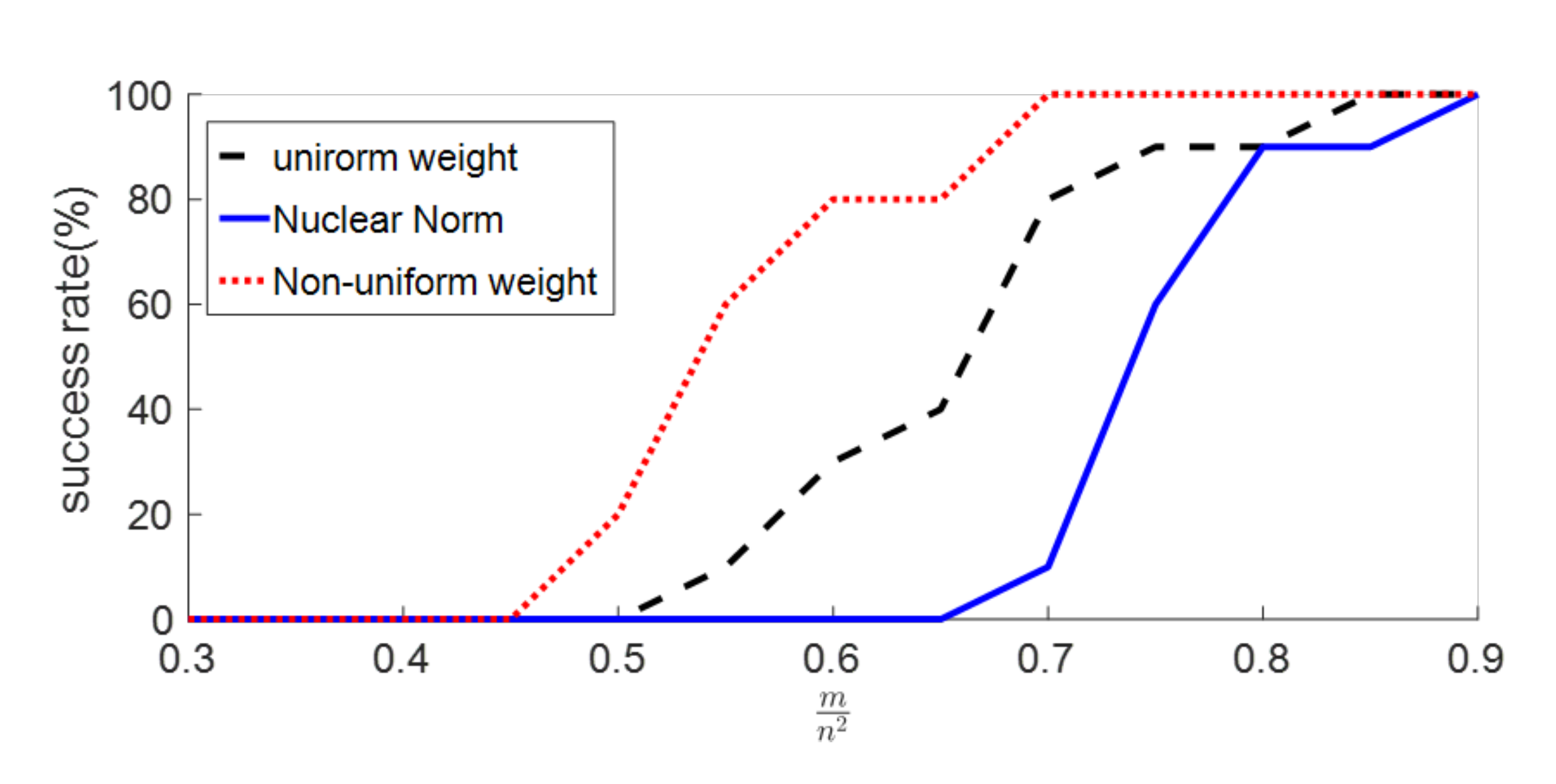}
	\includegraphics[width=3.5in,height=1.5in]{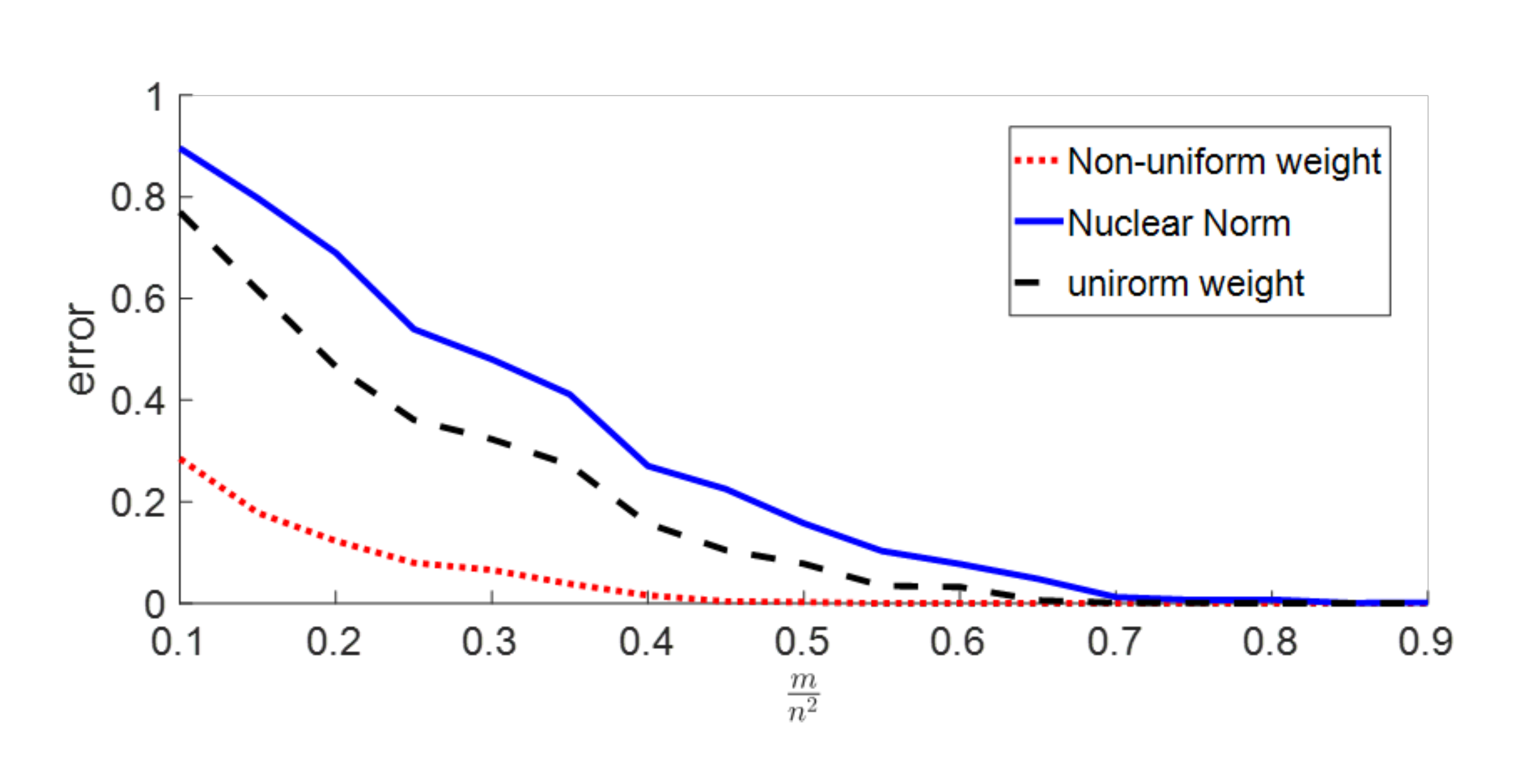}
	\caption{Matrix completion using different approaches in noise-less case. Principal angels are considered as $\bm\theta_u = [2.01,8.28,15.55,20.26]$ and $\bm\theta_v = [2.09,10.5,19.45,22.00]$.}
 
	\label{Fig1com}
\end{figure*}
\begin{figure*}
	%	\hspace*{-.5cm}
	%	\centering
	\includegraphics[width=3.5in,height=1.5in]{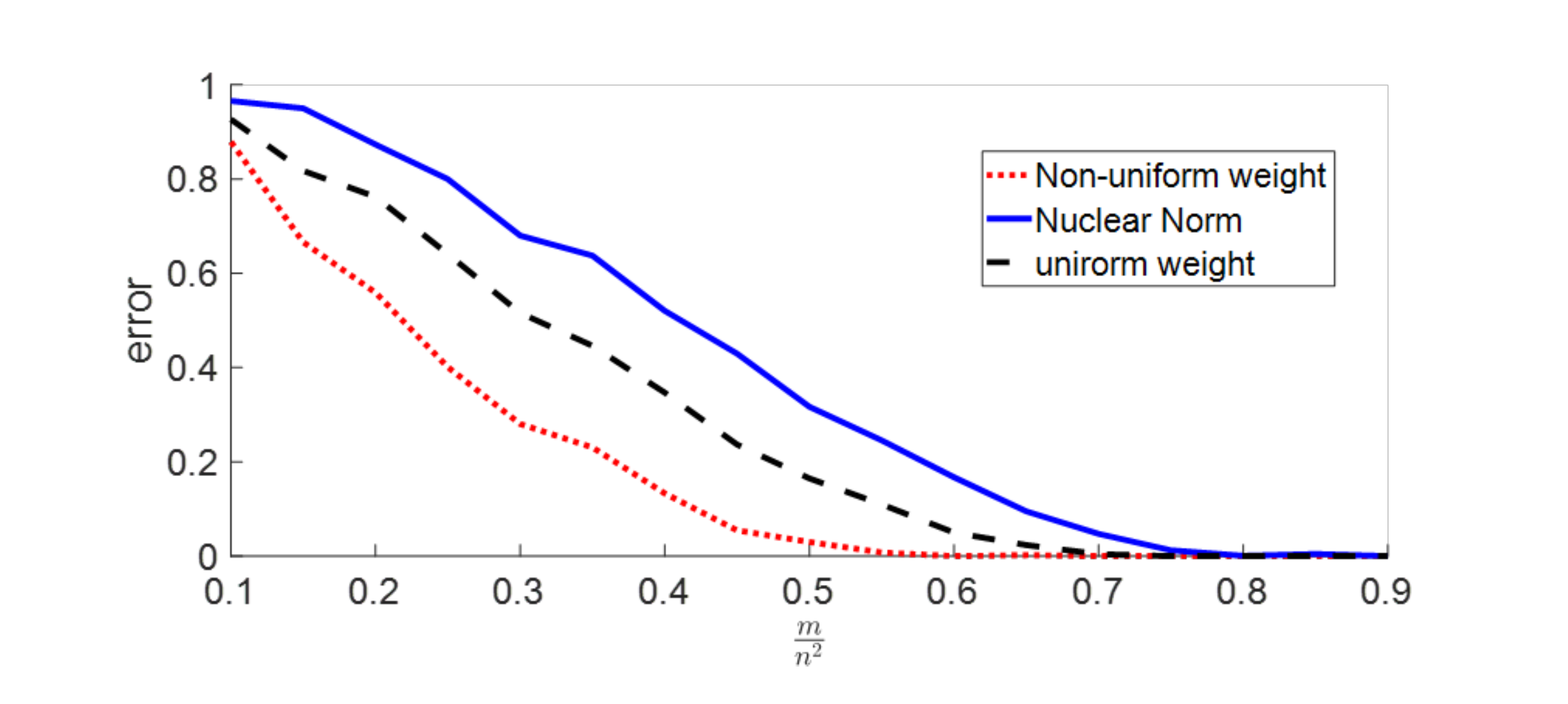}
	\includegraphics[width=3.5in,height=1.5in]{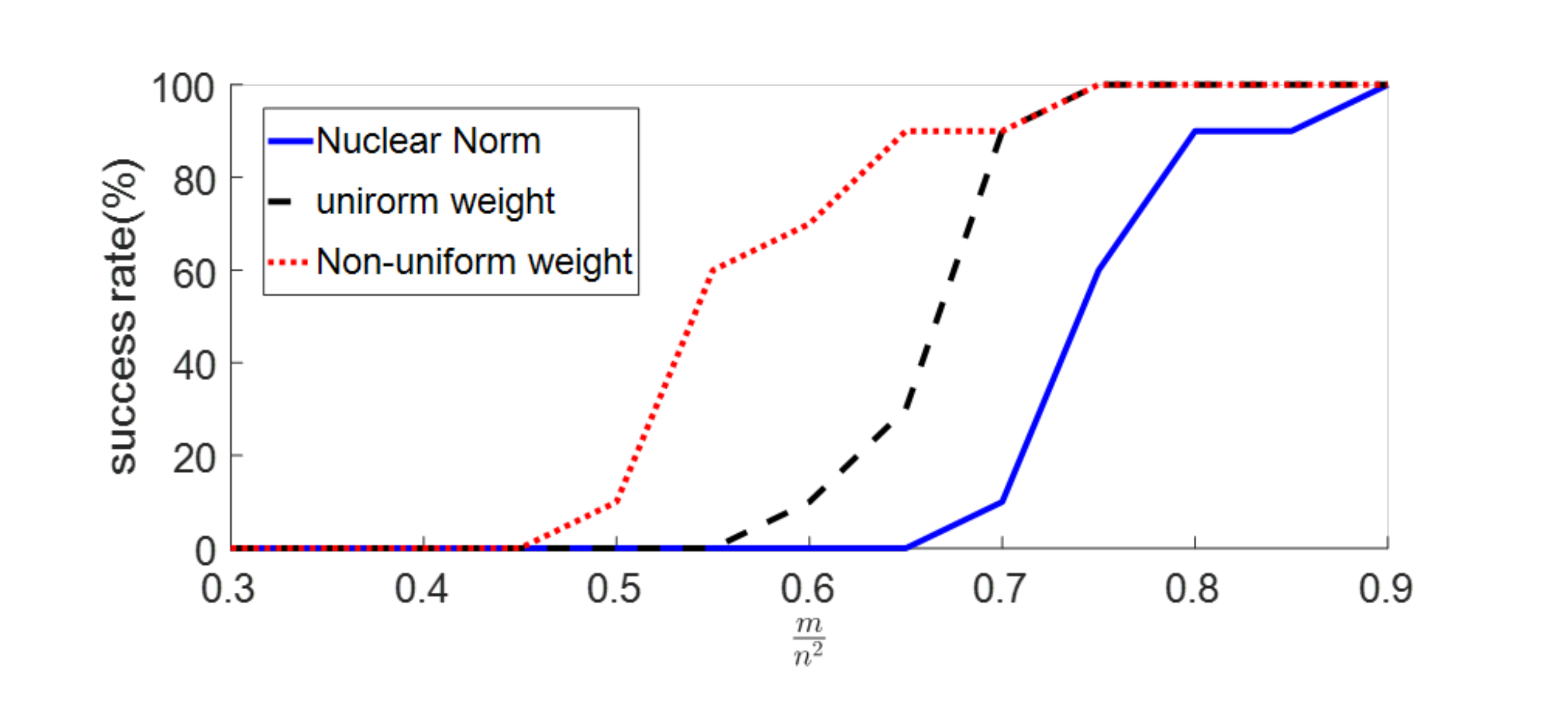}
	\caption{Matrix completion using different approaches in noise-less case. Principal angels are considered as $\bm\theta_u = [1.32,1.72,2.11,3.07]$ and $\bm\theta_v=[1.08,1.70,2.37,2.73]$.}
	
	\label{Fig2com}
\end{figure*}
\begin{figure*}
	%	\hspace*{-.5cm}
	%	\centering
	\includegraphics[width=3.5in,height=1.5in]{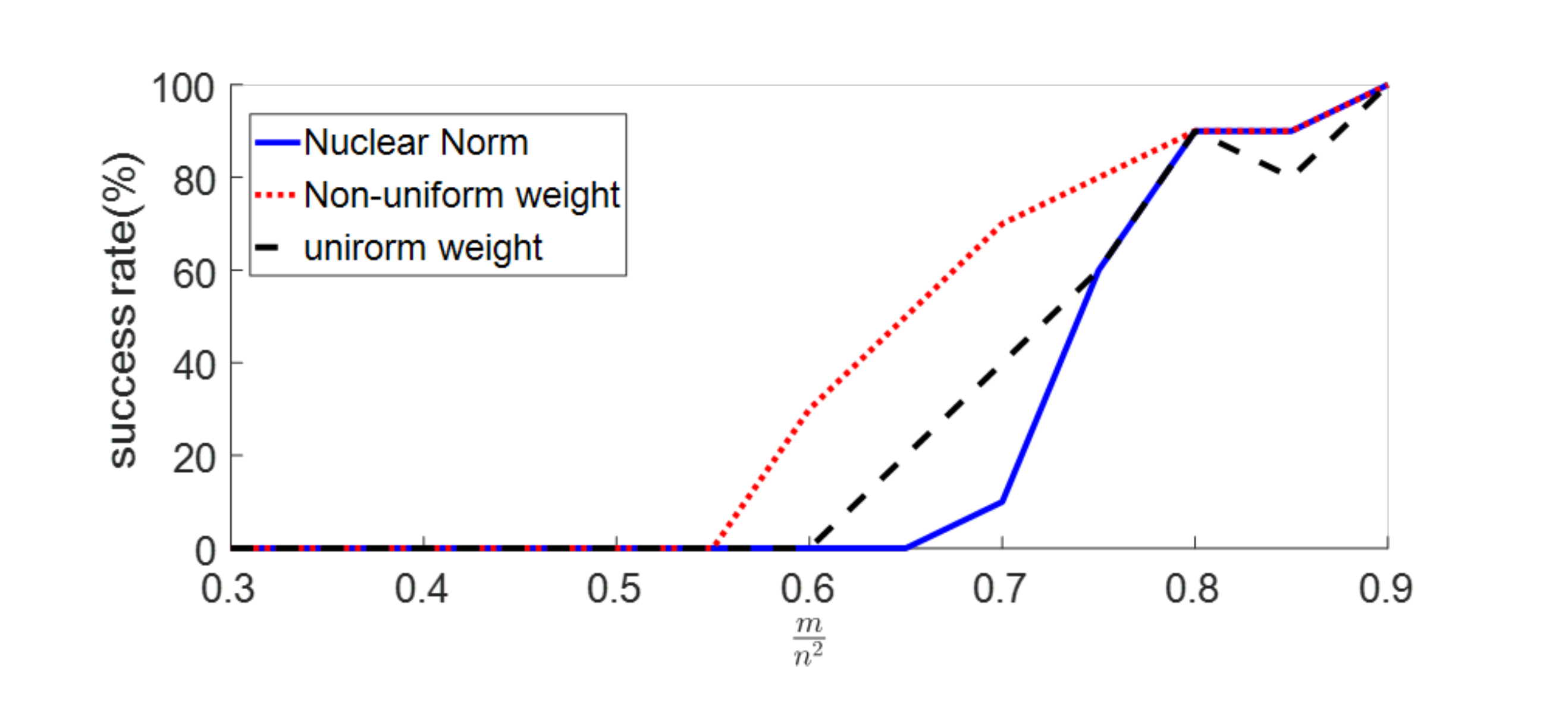}
	\includegraphics[width=3.5in,height=1.5in]{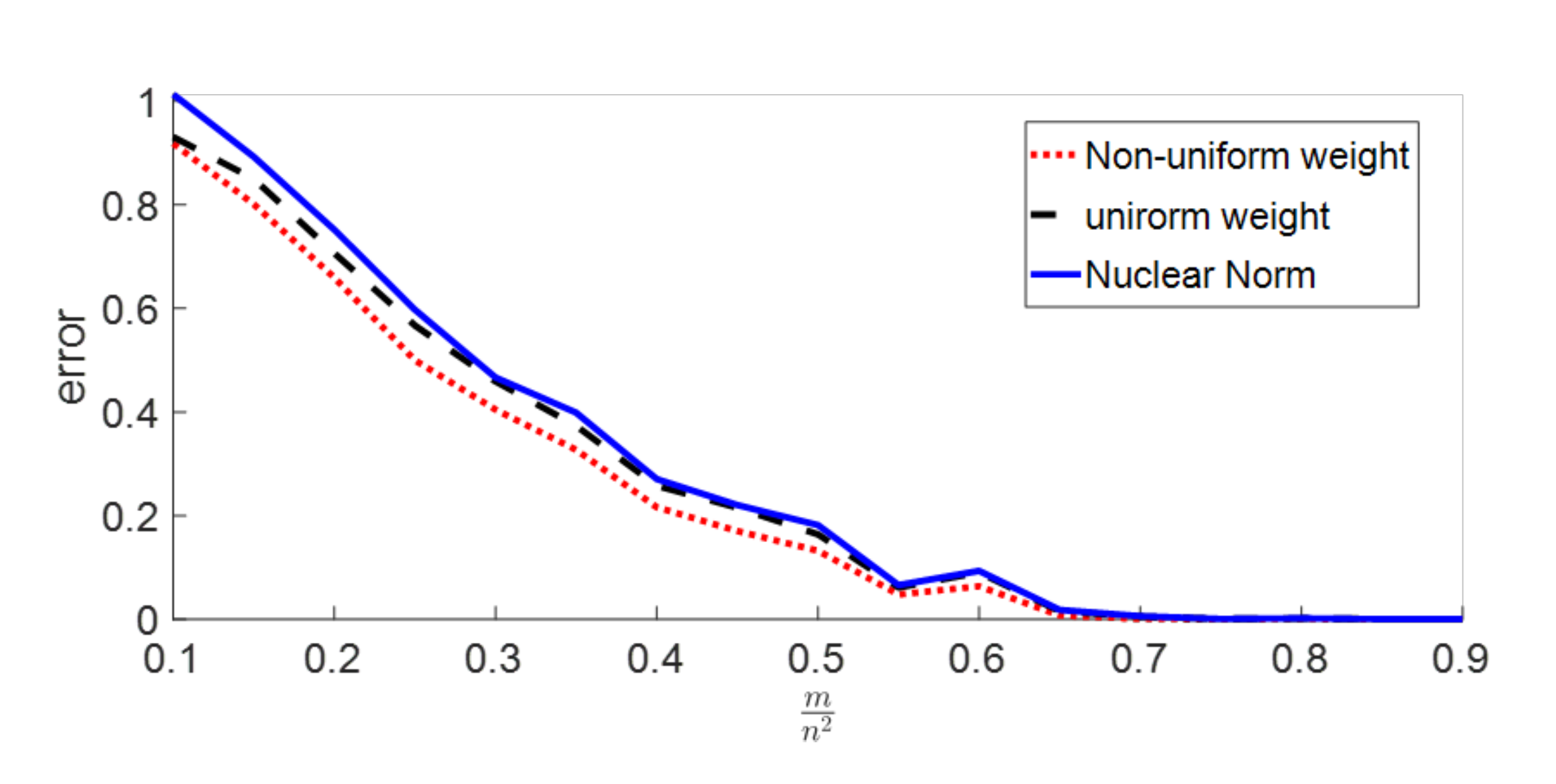}
	\caption{Matrix completion using different approaches in noise-less case. Principal angels are considered as $\bm\theta_u = [1.32,1.72,2.11,3.07]$ and $\bm\theta_v=[1.08,1.70,2.37,2.73]$.}
	
	\label{Fig3com}
\end{figure*}
In this section, we provide simulation experiments to represent that non-uniform weighting strategy performs better than uniform weighting approach in matrix completion problem. The simulation results are generated using CVX package and optimal weights are obtained by numerical optimization. Consider a square matrix $\bm X \in \mathbb{R}^{ n \times n}$ with $n=20$ and rank $r=4$. We use a perturbation matrix $\bm N \in \mathbb{R}^{n \times n}$, with independent Gaussian random elements of mean zero and variance $10^{-4}$ to have $\bm{X}'=\bm X + \bm N$. Then, we construct the prior subspaces $\bm{\widetilde{\mathcal{U}}}_{r'=r+4}$ and $\bm{\widetilde{\mathcal{V}}}_{r'=r+4}$ as spans of $\bm X'$ and $\bm{X}'^H$, respectively. Also $\bm{\theta}_u \in [0,90]^r$ and $\bm{\theta}_v \in [0,90]^r$ are the known principal angels between subspaces $[\bm{{\mathcal{U}}}_{r} , \bm{\widetilde{\mathcal{U}}}_{r'}]$ and $[\bm{{\mathcal{V}}}_{r} , \bm{\widetilde{\mathcal{V}}}_{r'}]$, respectively. The bases $\bm{{U}}_{r}$ and $\bm{{U}}_{r'}$ without loss of generality can be defined in such a way that 
\begin{align}
\bm{{U}}_{r}^H\bm{\widetilde{U}}_{r'}=[{\rm cos}\bm \theta_u~~\bm{0}_{r \times r'-r}], \quad 
\bm{{V}}_{r}^H\bm{\widetilde{V}}_{r'}=[{\rm cos}\bm \theta_v~~\bm{0}_{r \times r'-r}]
\end{align}
i.e. $\bm{{U}}_{r}$ and $\bm{\widetilde{U}}_{r'}$ can be considered as left and right singular matrices of $\bm{{U}}_{r}^H\bm{\widetilde{U}}_{r'}$. Similar definitions are also done for $\bm{{V}}_{r}$ and $\bm{\widetilde{V}}_{r'}$.

In this section, we compare the performance of the standard matrix completion \eqref{nuclear_prob} with the strategy of non-uniform weights  \eqref{prior_nuclear_prob}. Also the standard nuclear norm minimization with optimal weights in different $\bm\theta_u$ and $\bm\theta_v$ are compared. Each experiment is repeated $50$ times with different sampled entries and noise (in noisy cases). Considering $\hat{\bm X}$ as the solution of the problem, the normalized recovery error (NRE) is defined as: ${\rm NRE}:=\frac{\| \hat{\bm X} - \bm X \|_F}{\| \bm X \|_F} \nonumber$. NRE less than $10^{-4}$ shows a successful experiment.

Success rate and NRE (in the noiseless case) are shown in Fig. \ref{Fig1com}. In this experiment, prior information is assumed to have good accuracy, and the principal angles between subspaces are considered as $\bm\theta_u = [1.32,1.72,2.11,3.07]$ and $\bm\theta_v=[1.08,1.70,2.37,2.73]$ degrees. As it can be observed, matrix completion with non-uniform weights outperforms the standard problems and unweighted algorithm. 

In Fig. \ref{Fig2com}, we investigate a case with similar parameters assuming $\bm\theta_u = [2.01,8.28,15.55,20.26]$ and $\bm\theta_v = [2.09,10.5,19.45,22.00]$ with the difference that some directions are accurate and some are not. As expected, we see that matrix completion with non-uniform weights performs better than other methods.

In Fig. \ref{Fig3com}, we consider prior information with weak accuracy, i.e. $\bm\theta_u = [40.87,49.63,50.55,69.39]$ and $\bm\theta_v = [28.76,37.83,40.52,63.65]$. In this experiment, similar to previous ones, it can be again observed that non-uniformly weighted matrix completion has superior performance compared to the other methods.
\section{Conclusion}\label{sec.conclusion}
In this work, we designed a novel framework in order to exploit prior subspace information in matrix competition problem. We first developed a weighted optimization problem to promote both prior knowledge and low-rank feature. Then, we proposed a novel theoretical way to obtain the optimal weights by minimizing the required observations. Numerical results were also provided which demonstrate the superior accuracy of our proposed approach compared to the state-of-the-art methods.

\begin{appendices}
\section{Required Lemmas and Background}\label{sec.required_lemmas}
In this section, some essential lemmas are provided which are required in the proof of Theorem \eqref{theory_main}.
\subsection{Constructing the Bases}
This section introduces the bases in order to simplify the proofs. 
\begin{lem} \label{Lemma 3-6-1} \cite{daei2018optimal} Consider $\bm{X}_r \in \mathbb{R}^{n \times n}$ as a rank $r$ matrix with column and row subspaces $\bm{\mathcal{U}}_r$ and $\bm{\mathcal{V}}_r$, respectively. Also, let $\widetilde{\bm{\mathcal{U}}}_{r'}$ and $\widetilde{\bm{\mathcal{V}}}_{r'}$ of dimension $r'\geq r$ with $r$ known principal angles $\bm{\theta}_u$ and $\bm{\theta}_v$ with subspaces $\bm{\mathcal{U}}_r$ and $\bm{\mathcal{V}}$ as prior information. There exists orthogonal matrices $\bm{U}_r , \bm{V}_r \in \mathbb{R}^{n \times r}$ and $\widetilde{\bm{\mathcal{U}}}_{r'},\widetilde{\bm{\mathcal{V}}}_{r'} \in \mathbb{R}^{n \times r'}$ and $\bm{B}_L,\bm{B}_R \in \mathbb{R}^{n \times n}$ such that:
	\begin{align}
	&\bm{\mathcal{U}}_r = {\rm span}(\bm{U}_r), \quad \bm{\mathcal{V}}_r = {\rm span}(\bm{V}_r) \nonumber \\
	&\widetilde{\bm{\mathcal{U}}}_{r'} = {\rm span}(\widetilde{\bm{U}}_{r'}), \quad \widetilde{\bm{\mathcal{V}}}_{r'}  = {\rm span}(\widetilde{\bm{V}}_{r'})\nonumber \\
	&\bm{B}_L:=[\bm{U}_r~~\bm{U}_{1,r}'~~\bm{U}_{2,r'-r}' ~~\bm{U}_{n-r-r'}'']\in\mathbb{R}^{n\times n} \nonumber \\
	& \bm{B}_R:=[\bm{V}_r~~ \bm{V}_{1,r}' ~~\bm{V}_{2,r'-r}'~~ \bm{V}_{n-r-r'}'']\in\mathbb{R}^{n\times n}
	\end{align}
	For definitions of the submatrices, see \cite[Section VI.A]{daei2018optimal}.
\end{lem}
The following relation can be concluded from lemma \ref{Lemma 3-6-1}:
\begin{align}
\widetilde{\bm{U}}_{r'}=\bm{B}_L
\begin{bmatrix}
{\rm cos}\bm{\theta}_u &  \\ 
-{\rm sin}\bm{\theta}_u &  \\
& -\bm{I}_{r'-r} \\
&  \\
\end{bmatrix}
\end{align}
Therefore orthogonal projections onto the subspaces $\widetilde{\bm{\mathcal U}}_{r'}$ and $\widetilde{\bm{\mathcal U}}_{r'}^{\perp}$ are:
\begin{align}
&\bm{P}_{\widetilde{\bm{\mathcal U}}_{r'}}=\widetilde{\bm{U}}_{r'}\widetilde{\bm{U}}_{r'}^{H}=\nonumber \\
&\bm{B}_L 
\begin{bmatrix}
{\rm cos}^2\bm{\theta}_u & -{\rm sin}\bm{\theta}_u {\rm cos}\bm{\theta}_u &  &\\
-{\rm sin}\bm{\theta}_u {\rm cos}\bm{\theta}_u & {\rm sin}^2\bm{\theta}_u &  &\\
&  & \bm{I}_{r'-r} &\\
&&&
\end{bmatrix}
\bm{B}_L ^H \nonumber\\
&\bm{P}_{\widetilde{\bm{\mathcal U}}_{r'}^{\perp}}=\bm{I} - \bm{P}_{\widetilde{\bm{\mathcal U}}_{r'}}= \nonumber \\
&\bm{B}_L 
\begin{bmatrix}
{\rm sin}^2\bm{\theta}_u & {\rm sin}\bm{\theta}_u {\rm cos}\bm{\theta}_u &  &\\
{\rm sin}\bm{\theta}_u {\rm cos}\bm{\theta}_u & {\rm cos}^2\bm{\theta}_u &  &\\
&  & \bm{0}_{r'-r} &\\
&&&\bm{I}_{n-r'-r}
\end{bmatrix}\bm{B}_L ^H
\end{align}
We also have:
%\begin{align}
%\bm{Q}_{\widetilde{\bm{\mathcal U}}_{r'}}:=\widetilde{\bm{U}}_{r'} \bm\Lambda \widetilde{\bm{U}}_{r'}^H + \bm{P}_{\widetilde{\bm{\mathcal U}}^{\perp}}= \nonumber  
%\end{align}
\begin{align}\label{34}
&\bm{Q}_{\widetilde{\bm{\mathcal{U}}}_{r^{\prime}}} := \widetilde{\bm{U}}_{r^{\prime}}\bm{\Lambda}\widetilde{\bm{U}}_{r^{\prime}}^{\rm{H}} + \bm{P}_{\widetilde{\bm{\mathcal{U}}}^{\perp}} \nonumber \\
& = \bm{B}_L\left[\begin{array}{ccc}
\bm{\Lambda}_1 \cos^2 \mathbf{\bm{\theta}}_{u} + \sin^2 \mathbf{\bm{\theta}}_{u} \\
(\bm{I}-\bm{\Lambda}_1) \sin \mathbf{\bm{\theta}}_{u} \cos \mathbf{\bm{\theta}}_{u}\\
\\ \\
\end{array}\right.\nonumber\\
&\quad \quad \quad \quad \quad \quad \left.\begin{array}{ccc}
(\bm{I}-\bm{\Lambda}_1) \sin\mathbf{\bm{\theta}}_{u} \cos \mathbf{\bm{\theta}}_{u}&  &  \\
\bm{\Lambda}_1 \sin^2 \mathbf{\bm{\theta}}_{u} + \cos^2 \mathbf{\bm{\theta}}_{u} &  &  \\
&\bm{\Lambda}_2&  \\
&  &\bm{I}_{n-r^{\prime}-r}
\end{array}\right]\bm{B}_L^{\rm H},
\end{align}	
in which $ \bm{\Lambda} : = \begin{bmatrix}
\bm{\Lambda}_1 \in \mathbb{R}^{r \times r}& \\
& \bm{\Lambda}_2 \in \mathbb{R}^{r^{\prime}-r \times r^{\prime}-r}
\end{bmatrix} $.

Now in order to rewrite $\bm{Q}_{\widetilde{\bm{\mathcal{U}}}_{r^{\prime}}}$ to contain an upper-triangular matrix, we first define the orthonormal base:
\begin{multline}
\bm{O}_L := \left[
\begin{matrix}
(\bm{\Lambda}_1 \cos^2 \mathbf{\bm{\theta}}_{u} + \sin^2 \mathbf{\bm{\theta}}_{u}).\bm{\Delta}_L^{-1}  \\
-(\bm{I}-\bm{\Lambda}_1) \sin \mathbf{\bm{\theta}}_{u} \cos \mathbf{\bm{\theta}}_{u} .\bm{\Delta}_L^{-1}\\
\\ 
\\
\end{matrix}\right.               
\\
\left.
\begin{matrix}
-(\bm{I}-\bm{\Lambda}_1) \sin \mathbf{\bm{\theta}}_{u} \cos \mathbf{\bm{\theta}}_{u} .\bm{\Delta}_L^{-1} &  &   \\
(\bm{\Lambda}_1 \cos^2 \mathbf{\bm{\theta}}_{u} + \sin^2 \mathbf{\bm{\theta}}_{u}) .\bm{\Delta}_L^{-1}  &  &  \\ 
& \bm{I}_{r^{\prime}-r} &   \\
&  & \bm{I}_{n-r^{\prime}-r}
\end{matrix}\right],
\end{multline}
in which since $ \bm{\Lambda}_1 \succeq \bm{0} $, $ \bm{\Delta}_L := \sqrt{\bm{\Lambda}_1^2 \cos^2 \mathbf{\bm{\theta}}_{u} + \sin^2 \mathbf{\bm{\theta}}_{u}} \in \mathbb{R}^{n \times n }$  is an invertible matrix . Now \eqref{34} can be rewritten as:
\begin{align}
&\bm{Q}_{\widetilde{\bm{\mathcal{U}}}_{r^{\prime}}} =\bm{B}_L (\bm{O}_L\bm{O}_L^{\rm H})
\left[\begin{array}{ccc}
\bm{\Lambda}_1 \cos^2 \mathbf{\bm{\theta}}_{u} + \sin^2 \mathbf{\bm{\theta}}_{u}\\
(\bm{I}-\bm{\Lambda}_1) \sin \mathbf{\bm{\theta}}_{u} \cos \mathbf{\bm{\theta}}_{u}\\
\\
\\
\end{array}\right.\nonumber\\
&\quad \quad \quad \left.\begin{array}{ccc}
(\bm{I}-\bm{\Lambda}_1) \sin\mathbf{\bm{\theta}}_{u} \cos \mathbf{\bm{\theta}}_{u}& & \\
\bm{\Lambda}_1 \sin^2 \mathbf{\bm{\theta}}_{u} + \cos^2 \mathbf{\bm{\theta}}_{u} & & \\
&\bm{\Lambda}_2 & \\
& &\bm{I}_{n-r^{\prime}-r}
\end{array}\right]\bm{B}_L^{\rm H} \nonumber \\
&=\bm{B}_L \bm{O}_L \left[\begin{array}{ccc}
\bm{\Delta}_L \\
\\
\\
\\
\end{array}\right.\nonumber\\
&\left.\begin{array}{ccc}
(\bm{I}-\bm{\Lambda}_{1}^{2}) \sin \mathbf{\bm{\theta}}_{u} \cos \mathbf{\bm{\theta}}_{u}.\bm{\Delta}_L^{-1}& & \\
\bm{\Lambda}_1\bm{\Delta}_L^{-1} & &  \\
&\bm{\Lambda}_2&  \\
&  &\bm I_{n-r'-r}
\end{array}\right]\bm{B}_L^{\rm H} \nonumber \\
& =: \bm{B}_L \bm{O}_L \begin{bmatrix}
\bm{L}_{11} & \bm{L}_{12} & & \\
& \bm{L}_{22} & & \\ 
&‌&\bm{\Lambda}_2‌& \\
& & &\bm{I}_{n-r^{\prime}-r}
\end{bmatrix} \bm{B}_L^{\rm H}  \nonumber \\
& = \bm{B}_L \bm{O}_L \bm{L} \bm{B}_L^{\rm H}, 
\end{align}
where $\bm{L} \in \mathbb{R}^{n \times n }$ is  a block upper-triangular matrix:
\begin{align}\label{37}
&\bm{L} :=\begin{bmatrix}
\bm{L}_{11} & \bm{L}_{12} & & \\
& \bm{L}_{22} & & \\ 
&‌&\bm{\Lambda}_2‌& \\
& & &\bm{I}_{n-r^{\prime}-r}
\end{bmatrix}  \nonumber \\
&=\begin{bmatrix}
\bm{\Delta}_L  &  (\bm{I}-\bm{\Lambda}_{1}^{2}) \sin \mathbf{\bm{\theta}}_{u} \cos \mathbf{\bm{\theta}}_{u}.\bm{\Delta}_L^{-1}  & & \\
& \bm{\Lambda}_1\bm{\Delta}_L^{-1} &  & \\
& & \bm{\Lambda}_2& \\
& & &\bm{I}_{n-r^{\prime}-r} \\
\end{bmatrix}.
\end{align}
Since matrices $\bm{B}_L$ and $\bm{O}_L$ indicate orthonormal bases, it follows that $\| \bm{Q}_{\widetilde{\bm{\mathcal{U}}}_{r^{\prime}}} \| = \| \bm{L} \| = 1$.
%\begin	{align}\label{38}
%\| \bm{Q}_{\widetilde{\bm{\mathcal{U}}}_{r^{\prime}}} \| = \| \bm{L} \| = 1 .
%\end{align} 
Similar results can  also be deduced for the row subspace:
\begin{align}
&\bm{R} :=\begin{bmatrix}
\bm{R}_{11} & \bm{R}_{12} & & \\
& \bm{R}_{22} & & \\
& &\bm{\Gamma}_2 & \\
& & & \bm{I}_{n-r^{\prime}-r}
\end{bmatrix}  \nonumber \\
&= \begin{bmatrix}
\bm{\Delta}_R  &  (\bm{I}-\bm{\Gamma}_{1}^{2}) \sin \mathbf{\bm{\theta}}_{v} \cos \mathbf{\bm{\theta}}_{v}.\bm{\Delta}_R^{-1}  & &  \\
& \bm{\Gamma}_{1}\bm{\Delta}_R^{-1} & & \\
& & \bm{\Gamma}_2 & \\
& & & \bm{I}_{n-r^{\prime}-r}  
\end{bmatrix},
\end{align}
where $ \bm{\Delta}_R := \sqrt{\bm{\Gamma}_{1}^{2} \cos^2 \mathbf{\bm{\theta}}_{v} + \sin^2 \mathbf{\bm{\theta}}_{v}} $ and  $ \bm{\Delta}_L $ have similar properties. Considering $ \bm{H} \in \mathbb{R}^{n \times n } $  as an arbitrary matrix, one can say:
\begin{align}\label{40}
&\bm{Q}_{\widetilde{\bm{\mathcal{U}}}_{r^{\prime}}}\bm{H}\bm{Q}_{\widetilde{\bm{\mathcal{V}}}_{r^{\prime}}} = \bm{B}_L \bm{O}_L \bm{L} (\bm{B}_L^{\rm H} \bm{H} \bm{B}_R) \bm{R}^{\rm H} \bm{O}_R^{\rm H}  \bm{B}_R^{\rm H}   \nonumber \\
&= \bm{B}_L \bm{O}_L \bm{L} \overline{\bm{H}} \bm{R}^{\rm H} \bm{O}_R^{\rm H} \bm{B}_R^{\rm H} \quad (\overline{\bm{H}}:=\bm{B}_L^{\rm H} \bm{H} \bm{B}_R )  \nonumber \\
&=:\bm{B}_L \bm{O}_L \bm{L}
\begin{bmatrix}
\overline{\bm{H}}_{11} & \overline{\bm{H}}_{12} &‌\overline{\bm{H}}_{13} & \overline{\bm{H}}_{14}\\
\overline{\bm{H}}_{21} & \overline{\bm{H}}_{22} & \overline{\bm{H}}_{23} & \overline{\bm{H}}_{24}\\
\overline{\bm{H}}_{31} & \overline{\bm{H}}_{32} &‌\overline{\bm{H}}_{33} & \overline{\bm{H}}_{34}\\
\overline{\bm{H}}_{41} &  \overline{\bm{H}}_{42} & \overline{\bm{H}}_{43}& \overline{\bm{H}}_{44}
\end{bmatrix}\bm{R}^{\rm H} \bm{O}_R^{\rm H} \bm{B}_R^{\rm H}.
\end{align}
Since ${\rm span}(\bm{X}_r) = {\rm span}(\bm{U}_r)$ and ${\rm span}(\bm{X}_r^{\rm H}) = {\rm span}(\bm{V}_r)$ and with upper triangular matrices $\bm{L}$ and $\bm{R}$, we can rewrite $ \bm{Q}_{\widetilde{\bm{\mathcal{U}}}_{r^{\prime}}}  \bm{X}_r \bm{Q}_{\widetilde{\bm{\mathcal{V}}}_{r^{\prime}}} $ in terms of new bases:
\begin{align}\label{42}
&\bm{Q}_{\widetilde{\bm{\mathcal{U}}}_{r^{\prime}}}  \bm{X}_r \bm{Q}_{\widetilde{\bm{\mathcal{V}}}_{r^{\prime}}} = \bm{B}_L \bm{O}_L \bm{L} (\bm{B}_L^{\rm H} \bm{X}_r \bm{B}_R) \bm{R}^{\rm H} \bm{O}_R^{\rm H}  \bm{B}_R^{\rm H}   \nonumber \\
&= \bm{B}_L \bm{O}_L \bm{L} \overline{\bm{X}}_r \bm{R}^{\rm H} \bm{O}_R^{\rm H} \bm{B}_R^{\rm H} \quad (\overline{\bm{X}}_r:=\bm{B}_L^{\rm H} \bm{X}_r \bm{B}_R )  \nonumber \\
&=:\bm{B}_L \bm{O}_L \bm{L}
\begin{bmatrix} 
\overline{\bm{X}}_{r,11} &  \\
&  \bm{0}_{n-r} 
\end{bmatrix}\bm{R}^{\rm H} \bm{O}_R^{\rm H} \bm{B}_R^{\rm H} \nonumber \\ 
&=\bm{B}_L \bm{O}_L 
\begin{bmatrix}
\bm{L}_{11}\overline{\bm{X}}_{r,11}\bm{R}_{11} &  \\
&  \bm{0}_{n-r} 
\end{bmatrix}\bm{O}_R^{\rm H} \bm{B}_R^{\rm H}.
\end{align}
\begin{lem}\label{lem 4}
	The operator norms regarding the sub-blocks of L in \eqref{37} are as follows:
	\begin{align}
	&\|\bm{L}_{11}\| = \|\bm{\Delta}_L\| = \max_{i} \sqrt{ \lambda_{1}^{2}(i)\cos^2 \mathbf{\theta}_{u}(i) + \sin^2 \mathbf{\theta}_{u}(i)}, \nonumber \\ 
	& \|\bm{L}_{12}\| = \max_i \sqrt{\frac{(1-\lambda_{1}^{2}(i))^2 \cos^2 \mathbf{\theta}_{u}(i) \sin^2 u_i}{\lambda_{1}^{2}(i)\cos^2 \mathbf{\theta}_{u}(i) + \sin^2 \mathbf{\theta}_{u}(i)}},  \nonumber \\
	& \|\bm{I}_r - \bm{L}_{22} \| = \max_i \frac{\lambda_{1}(i)-\sqrt{\lambda_{1}^{2}(i)\cos^2 \mathbf{\theta}_{u}(i) + \sin^2 \mathbf{\theta}_{u}(i)}}{\sqrt{\lambda_{1}^{2}(i)\cos^2\mathbf{\theta}_{u}(i) + \sin^2 \mathbf{\theta}_{u}(i)}},  \nonumber \\
	& \label{eq:L11L12}\|[\bm{L}_{11} \quad  \bm{L}_{12}]\| = \max_i  \sqrt{  \frac{\lambda_{1}^{4}(i) \cos^2 \mathbf{\theta}_{u}(i) + \sin^2 \mathbf{\theta}_{u}(i)}{\lambda_{1}^{2}(i) \cos^2 \mathbf{\theta}_{u}(i) + \sin^2 \mathbf{\theta}_{u}(i) } } \\ 
	& \label{eq:L'} \|\bm L ^ \prime \|^{2} = \max_i {d}_i(\bm{\theta}_{u},\bm \lambda_{1},\bm \lambda_{2}) \\
	& \Big\|\begin{bmatrix}
	\bm{I}_{r} - \bm{L}_{22} & \\
	& \bm{I}_{r} - \bm{\Lambda}_{2}
	\end{bmatrix} \Big\| =  \max ~ \Big \{  \max_i (1-\lambda_{2}(i))\nonumber \\
	&‌ \quad \quad \quad \quad \quad \quad , \max_i \Big(1 - \frac{\lambda_{1}(i)}{\sqrt{\lambda_{1}^{2}(i)\cos^2 \mathbf{\theta}_{u}(i) + \sin^2 \mathbf{\theta}_{u}(i)}}\Big) \Big \}, 
	\end{align} 
	where $d_i$ is defined as:% in \eqref{eq:d1} and \eqref{eq:d2}. 
	\begin{align}
	&d_1(\bm\theta,\bm a,\bm b):= \underset{i}{\rm max} \big( \big( \frac{a(i)}{sqrt{a^2(i)\cos^\theta(i)+\sin^2\theta(i)}} -1 \big)^2 \nonumber \\
	&+ \frac{(1-a(i))^2\cos^2\theta(i)\sin^2\theta(i)}{a(i)^2\cos^2\theta(i)\sin^2\theta(i)} \big) \nonumber \\
	& d_2(\bm\theta,\bm a,\bm b):=\underset{i}{\rm max}(b(i)-1)^2
	\end{align}
	The same  equalities hold for sub-blocks of $ \bm{R} $. 
	
	Proof.  See Appendix \ref{proof lemma 4}.
\end{lem}
\subsection{Support Definitions}
Let $\bm{X}_r \in \mathbb{R}^{n \times n}$ be a rank-$r$ matrix which is obtained via the truncated SVD of $ \bm{X} $: 
$$ \bm{X} = \bm{X}_r + \bm{X}_{r^+} = \bm{U}_r\overline{\bm{X}}_{r,11}\bm{V}_r^{\rm H} +‌ \bm{X}_{r^+},$$
where $ \bm{U}_r $ and $ \bm{V}_r $ are some orthogonal bases of column and row spaces of $\bm{X}_r$, and thus $ \overline{\bm{X}}_{r,11} $ is not necessarily diagonal. Also consider that  $ \bm{\mathcal{U}}_{r} = {\rm span}(\bm{U}_r)= {\rm span}(\bm{X}_r) $ and $ \bm{\mathcal{V}}_{r} = {\rm span}(\bm{V}_r)= {\rm span}(\bm{X}_r^{\rm H}) $ are column and row subspaces of $  \bm{X}_r$, respectively. Then the support of $ \bm{X}_r $ can be defined as:  
\begin{align}\label{44}
&\bm{T} := \{\bm{Z}\in\mathbb{R}^{n\times n} : \bm{Z}=\bm{P}_{\bm{\mathcal{U}}_r}\bm{Z}\bm{P}_{\bm{\mathcal{V}}_r}+\bm{P}_{\bm{\mathcal{U}}_r}\bm{Z}\bm{P}_{\bm{\mathcal{V}}_r^\perp} \nonumber \\
&\quad \quad \quad \quad \quad \quad 	+\bm{P}_{\bm{\mathcal{U}}_r^\perp}\bm{Z}\bm{P}_{\bm{\mathcal{V}}_r^\perp}\}   = {\rm supp}(\bm{X}_r),
\end{align}
and the orthogonal projection onto $\bm{T}$ and $\bm{T}^{\perp}$ can be defined as
\begin{align}
\mathcal{P}_{\bm{T}}(\bm{Z}) = \bm{P}_{\bm{\mathcal{U}}}\bm{Z} +‌\bm{Z}\bm{P}_{\bm{\mathcal{V}}} - \bm{P}_{\bm{\mathcal{U}}}\bm{Z}\bm{P}_{\bm{\mathcal{V}}},  \mathcal{P}_{\bm{T}^{\perp}}(\bm{Z}) =  \bm{P}_{\bm{\mathcal{U}}^{\perp}}\bm{Z}\bm{P}_{\bm{\mathcal{V}}^{\perp }}.
\end{align} 
We can rewrite $ \bm{T} $  using Lemma \ref{Lemma 3-6-1} as 
\begin{align} \label{3-42}
&\bm{T} = \Big\{\bm{Z}\in\mathbb{R}^{ n\times n} : \bm{Z}=\bm{B}_L \overline{\bm{Z}} \bm{B}_R^{\rm H}, \quad \overline{\bm{Z}}:=
\begin{bmatrix}
\overline{\bm{Z}}_{11} & \overline{\bm{Z}}_{12} \\
\overline{\bm{Z}}_{21} & \bm{0}_{n-r}
\end{bmatrix} \Big\}  \nonumber \\
&=\bm{B}_L \overline{\bm{T}} \bm{B}_R^{\rm H},
\end{align}
where $ \overline{\bm{T}} \subset \mathbb{R}^{n \times n} $  is the support of $ \overline{\bm{X}}_r = \bm{B}_L^{\rm H} \bm{X}_r \bm{B}_R $:
\begin{align}
\overline{\bm{T}} = \{\overline{\bm{Z}}\in\mathbb{R}^{ m\times n} : \overline{\bm{Z}}:=	\begin{bmatrix}
\overline{\bm{Z}}_{11} & \overline{\bm{Z}}_{12} \\
\overline{\bm{Z}}_{21} & \bm{0}_{n-r}
\end{bmatrix}\}.
\end{align}
For arbitrary 
\begin{align}
\overline{\bm{Z}}:=	\begin{bmatrix}
\overline{\bm{Z}}_{11} & \overline{\bm{Z}}_{12} \\
\overline{\bm{Z}}_{21} & \overline{\bm{Z}}_{22}
\end{bmatrix} \in \mathbb{R}^{n\times n},
\end{align}
the orthogonal projection onto $ \overline{\bm{T}} $ and its complement $ \overline{\bm{T}}^{\perp} $ are  
\begin{align}\label{47}
\mathcal{P}_{\overline{\bm{T}}}(\overline{\bm{Z}}) = \begin{bmatrix}
\overline{\bm{Z}}_{11} & \overline{\bm{Z}}_{12} \\
\overline{\bm{Z}}_{21} & \bm{0}_{n-r}
\end{bmatrix}, \quad \mathcal{P}_{\overline{\bm{T}}^{\perp}} (\overline{\bm{Z}})= \begin{bmatrix}
\bm{0}_{r} & \\
& \overline{\bm{Z}}_{22}
\end{bmatrix}, 
\end{align}
respectively. Since $ \bm{Z} = \bm{B}_L \overline{\bm{Z}} \bm{B}_R^{\rm H}  $, one can say:   
\begin{align}\label{48}
& \mathcal{P}_{\bm{T}}(\bm{Z}) = \bm{B}_L \mathcal{P}_{\overline{\bm{T}}}(\overline{\bm{Z}})\bm{B}_R^{H}, \quad
\mathcal{P}_{\bm{T}^{\perp}}(\bm{Z}) = \bm{B}_L \mathcal{P}_{\overline{\bm{T}}^{\perp}}(\overline{\bm{Z}}) \bm{B}_R^{H}.
\end{align}

\section{Proof of Theorem \ref{thm2}}\label{proof_th}
\begin{proof} 
For matrix $\bm{X}_r$ with rank $r$, If $\bm{U}_r$ and $\bm{V}_r$ indicate the orthogonal bases, then column and row subspaces are:
$$\bm{\mathcal{U}}_r = {\rm span}(\bm{X}_r) = {\rm span}(\bm{U}_r) , \bm{\mathcal{V}}_r = {\rm span}(\bm{X}_r^H) = {\rm span}(\bm{V}_r)$$
and for coherency of the $i$th row and $j$th column of $\bm{X}_r$  we have
\begin{align}
&\mu_i = \mu_i(\bm{\mathcal{U}}_r) := \frac{n}{r} \| \bm{U}_r[i,:] \|_{2}^2  \quad i \in [1:n]  \nonumber \\
&\nu_j = \nu_j(\bm{\mathcal{V}}_r) := \frac{n}{r} \| \bm{V}_r[j,:] \|_{2}^2  \quad j \in [1:n]
\end{align}	
As we can see in these definitions, coherency of subspaces are independent from selection of orthogonal bases. According to Definition \ref{coherency} and \eqref{max_mu}, the coherency of a matrix is equal to the maximum coherency of subspaces.
\begin{align}
\eta(\bm{X}_r) = \max_{i} \mu_i(\bm{\mathcal{U}}_r) \vee \max_{j} \nu_j(\bm{\mathcal{V}}_r) \label{max_mu}
\end{align} 
similarly, for subspaces $\breve{\bm{\mathcal{U}}}_r={\rm span}([\bm U , \bm{\widetilde{U}}_{r'}])$ and $\breve{\bm{\mathcal{V}}}_r={\rm span}([\bm V , \bm{\widetilde{V}}_{r'}])$ we have:
\begin{align}
\breve{\mu}_i = \breve{\mu}_i(\breve{\bm{\mathcal{U}}}_r)  \quad i \in [1:n] \quad
\breve{\nu}_j = \nu_j(\breve{\bm{\mathcal{V}}}_r)  \quad j \in [1:n]
\end{align}
For simplicity and in order to use coherency between subspaces, we define a diagonal matrix as:
\begin{align}
\mu=  \begin{bmatrix}
\mu_{1} & & \\
& \ddots & \\
& & \mu_{n}
\end{bmatrix}
\end{align}
although only $\mu_i$ is considered in this definition, but it can be expanded for other coherencies. \\
let $\| \bm A \|_{2 \rightarrow \infty }$ be the maximum $\ell_2$-norm of rows in $\bm A$, then
\begin{align}
&\|(\frac{\mu r}{n})^{\frac{-1}{2}} \bm{U}_r \|_{2 \rightarrow \infty} =1, \|(\frac{\nu r}{n})^{\frac{-1}{2}} \bm{V}_r \|_{2 \rightarrow \infty} =1  \nonumber \\
&\|(\frac{\mu r}{n})^{\frac{-1}{2}} \bm{U}^{\prime}_{r^{\prime}} \|_{2 \rightarrow \infty} \le \sqrt{\frac{r^{\prime}+r}{r}\max_{i}\frac{\breve{\mu}_i}{\mu_i}}  \nonumber \\
&\|(\frac{\nu r}{n})^{\frac{-1}{2}} \bm{V}^{\prime}_{r^{\prime}} \|_{2 \rightarrow \infty} \le \sqrt{\frac{r^{\prime}+r}{r}\max_{j}\frac{\breve{\nu}_j}{\nu_j}}
\end{align}
considering $\breve{\bm U}$ as the orthogonal bases of $\breve{\bm{\mathcal{U}}}$, correctness of above equations can be provided by:
\begin{align}
&\|(\frac{\mu r}{n})^{\frac{-1}{2}} \bm{U}^{\prime}_r \|_{2 \rightarrow \infty} = \max_{i} \frac{\| \bm{U}^{\prime}_r[i,:]\|_{2}}{\| \bm{U}_r[i,:]\|_{2}}  \nonumber \\
&\le \max_{i}\frac{\| \breve{\bm{U}_r}[i,:]\|_{2}}{\| \bm{U}_r[i,:]\|_{2}} = \sqrt{\max_{i}\frac{\breve{\mu}_i .{\rm dim}(\breve{\bm{U}_r})}{\mu_i r}}  \nonumber \\
& \le \sqrt{\frac{r^{\prime}+r}{r} \max_{i}\frac{\breve{\mu}_i}{\mu_i}} \quad ({\rm dim}(\breve{\bm{U}_r}) \le r^{\prime}+r)  \label{proof5}
\end{align} 
where the third inequality of \eqref{proof5} is followed by $\bm{\mathcal{U}_{r'}'}\subset \breve{\bm{\mathcal{U}}}$.\\
According to the definition of sampling operator in \eqref{sampling_opr}, we can say:
\begin{align}
\| \mathcal{R} _\Omega(\bm{X}) \|_{F} = \sum_{i,j = 1}^{n} \frac{\epsilon_{ij}}{p_{ij}} \bm{X}[i,j].\bm{C}_{i,j}	
\end{align}
Now, we provide some properties of sampling operator, which is imperative for developing our main Theory. In order to list these properties, at first we define following norms. Assume $\mu(\infty)$ and $\mu(\infty,2)$ measure the weighted $\ell_{\infty}$-norm and the maximum weighted $\ell_2$-norm of rows of a matrix, respectively. For a square matrix $\bm Z \in \mathbb{R}^{n \times n}$ we set:
\begin{align}
&\| \bm{Z} \|_{\mu(\infty)} = \| (\frac{\mu r}{n})^{\frac{-1}{2}} \bm{Z} (\frac{\nu r}{n})^{\frac{-1}{2}} \| _{\infty}  =\max_{i,j}\sqrt{\frac{n}{\mu_i r}}|\bm{Z}[i,j]|\sqrt{\frac{n}{\nu_j r}}|
\end{align}

\begin{align}
&\| \bm{Z} \|_{\mu(\infty,2)} = \| (\frac{\mu r}{n})^{\frac{-1}{2}} \bm{Z}\|_{2 \rightarrow \infty} \vee \|(\frac{\nu r}{n})^{\frac{-1}{2}} \bm{Z}^{\rm H}\| _{2 \rightarrow \infty}  \nonumber \\
&=(\max_{i}\sqrt{\frac{n}{\mu_i r}}\|\bm{Z}[i,:]\|_2)\vee (\max_{j}\sqrt{\frac{n}{\nu_j r}} \|\bm{Z}[:,j]\|_2)
\end{align} 
\begin{lem} \label{Lemma1}
	\cite{chen2015completing} Considering the sampling operator defined in \eqref{sampling_opr}, also assuming $\bm T$ as support of $\bm X_r$ and $\mathcal{P}_T $ as the orthogonal projection onto the support, then 
$$ \|(\mathcal{P}_{\bm{T}} - \mathcal{P}_{\bm{T}}\circ \mathcal{R} _\Omega\circ\mathcal{P}_{\bm{T}})(\bm{Z}) \|_{F \rightarrow F} \le \frac{1}{2}$$
on condition that
$$ \frac{(\mu_i +‌\nu_j)r \log n}{n} \le p_{ij} \le 1 \quad \forall i,j \in [1:n].$$
\end{lem}

\begin{lem}
	\cite{chen2015completing} Considering the assumptions in Lemma \ref{Lemma1}, for every fix matrix $\bm Z$, we expect	
	$$ \|(\mathcal{I} - \mathcal{R} _\Omega)(\bm{Z}) \| \le \| \bm{Z} \|_{\mu(\infty)} + \| \bm{Z} \|_{\mu(\infty,2)} $$
	under the same condition of Lemma \ref{Lemma1}.
%	provided that
%$$ \frac{(\nu_j + \nu_j)r{\rm log}n}{n} \leq p_{ij} \leq 1 ~\forall i,j\in[1:n]. $$
\end{lem}

\begin{lem}
	\cite{chen2015completing} Considering the assumptions in Lemma \ref{Lemma1}, for every fix matrix $\bm Z$ that $\mathcal{P} (\bm Z) = \bm Z$, we expect
	\begin{align*}
	\|(\mathcal{P}_{\bm{T}} - \mathcal{P}_{\bm{T}}\circ \mathcal{R} _\Omega\circ\mathcal{P}_{\bm{T}})(\bm{Z}) \|_{\mu(\infty,2)}\le \frac{1}{2}\| \bm{Z} \|_{\mu(\infty,2)} + \frac{1}{2}\| \bm{Z} \|_{\mu(\infty)}
	\end{align*}
	under the same condition of Lemma \ref{Lemma1}.
\end{lem}
\begin{lem}
	\cite{chen2015completing} Considering the assumptions in Lemma \ref{Lemma1}, for a fix matrix $\bm Z \in \bm T$,
	$$ \|(\mathcal{P}_{\bm{T}} - \mathcal{P}_{\bm{T}}\circ \mathcal{R} _\Omega\circ\mathcal{P}_{\bm{T}})(\bm{Z}) \|_{\mu(\infty)} \le \frac{1}{2}\| \bm{Z} \|_{\mu(\infty)} $$
	In some cases,  we will use $\overline{\mathcal{R} _\Omega}$ instead of $\mathcal{R} _\Omega$ which is defined as:
	\begin{align}
	\overline{\mathcal{R} _\Omega}(\overline{\bm{Z}}) = \bm{B}_L^{\rm H} \mathcal{R} _\Omega( \bm{B}_L\overline{\bm{Z}} \bm{B}_R^{\rm H}) \bm{B}_R \label{A40}
	\end{align}
	In accordance with the sampling operator $\mathcal{R} _\Omega$, we define the orthogonal projection $\mathcal{P}_p(\bm Z) $ in order to project the input to the support of $\mathcal{R} _\Omega$:
	\begin{align}
 	&\mathcal{P}_p(\bm{Z}) = \sum_{i,j=1}^{n}\epsilon_{ij}\bm{Z}[i,j]\bm{C}_{ij}  
	\end{align} 
	so similar to $\overline{\mathcal{R} _\Omega}$ and $\mathcal{P}_p$, we have:
	\begin{align}
	&\overline{\mathcal{P}_p}(\overline{\bm{Z}}) = \bm{B}_L^{\rm H} \mathcal{P}_p( \bm{B}_L\overline{\bm{Z}} \bm{B}_R^{\rm H}) \bm{B}_R \label{A42} 
	\end{align}
\end{lem}
In the following we provide some principal features about the recent defined variables.
\begin{lem}
	For an arbitrary matrix $\bm Z$ and $\overline{\bm Z}\bm{Z}_L^H \bm Z \bm{B}_R$ and for operators ${\mathcal{P}_p} , \overline{\mathcal{P}_p} , \mathcal{R} _\Omega$ and $\overline{\mathcal{A}}_p$ we have:
	\begin{align}
	\langle \overline{\bm{Z}},\overline{\mathcal{R} _\Omega}(\overline{\bm{Z}})\rangle = \langle \bm{Z},\mathcal{R} _\Omega(\bm{Z})\rangle
	\end{align}
	\begin{align}
	\| \overline{\mathcal{R} _\Omega}(\overline{\bm{Z}})\|_{F} = \| \mathcal{R} _\Omega(\bm{Z})\|_{F}
	\end{align} 
	Also if $0 \leq l \leq h \leq1$ and $p_{i,j}\subset [l,h]$, then
	\begin{align}
	(\overline{\mathcal{R} _\Omega}\circ\overline{\mathcal{R} _\Omega})(.) \succeq (\overline{\mathcal{R} _\Omega})(.)
	\end{align} 
	\begin{align}
	\| \overline{\mathcal{R} _\Omega}(.) \|_{F \rightarrow F} =\| \mathcal{R} _\Omega(.) \|_{F \rightarrow F} \le l^{-1} 
	\end{align} 
	in the last two equations and for operators $\mathcal{A}(\cdot)$ and $\mathcal{B}(\cdot)$, $\mathcal{A}(\cdot) \succeq \mathcal{B}(\cdot)$ means that for every matrix $\bm Z$, $\langle \bm Z , \mathcal A (\bm Z) \rangle \succeq \langle \bm Z , \mathcal B (\bm Z) \rangle $, so finally
	\begin{align}
	\overline{\mathcal{P}_p}(\overline{\bm{Z}})\|_{F} \le h\|\overline{\mathcal{R} _\Omega}(\overline{\bm{Z}})\|_{F}
	\end{align}
\end{lem}
Now considering these lemmas, let's complete the proof of Theorem \eqref{theory_main}. 

Consider $\bm X = \bm{X}_r$ is a rank $r$ matrix and our observation is measured in a noisy environment. let $\hat{\bm X}$ and $\bm H := \hat{\bm X} - \bm X$ be the solution and the estimation error of \eqref{prior_nuclear_prob}. Then one can say:
\begin{align}
\| \bm{Q}_{\widetilde{\bm{\mathcal{U}}}_{r}} (\bm{X}+\bm{H}) \bm{Q}_{\widetilde{\bm{\mathcal{V}}}_{r}} \|_{*} \le  
\| \bm{Q}_{\widetilde{\bm{\mathcal{U}}}_{r}} \bm{X} \bm{Q}_{\widetilde{\bm{\mathcal{V}}}_{r}} \|_{*} \label{A48}
\end{align}
for the right side of \eqref{A48} we have:
\begin{align}
&\| \bm{Q}_{\widetilde{\bm{\mathcal{U}}}_{r}} \bm{X} \bm{Q}_{\widetilde{\bm{\mathcal{V}}}_{r}} \|_{*} = \| \bm{Q}_{\widetilde{\bm{\mathcal{U}}}_{r}} \bm{X}_r \bm{Q}_{\widetilde{\bm{\mathcal{V}}}_{r}} \|_{*}  \nonumber \\
& \le \| \bm{B}_L \bm{O}_L \bm{L}  \overline{\bm{X}}_{r} \bm{R}^{\rm H} \bm{O}_R^{\rm H}  \bm{B}_R^{\rm H} \|_{*} =  \| \bm{L}  \overline{\bm{X}}_{r} \bm{R}^{\rm H} \|_{*}  \nonumber \\
&= \Bigg\|
\begin{bmatrix}
\bm{L}_{11}\overline{\bm{X}}_{r,11}\bm{R}_{11}  & \\
 &‌\bm{0}_{n-r}‌
\end{bmatrix}
\Bigg\|_{*}
\end{align}
which is held since $\bm{B}_R, \bm{O}_R, \bm{B}_L$ and $\bm{O}_L$ are orthogonal. For the left side of \eqref{A48} we have:
\begin{align}
&\| \bm{Q}_{\widetilde{\bm{\mathcal{U}}}_{r}} (\bm{X}+\bm{H}) \bm{Q}_{\widetilde{\bm{\mathcal{V}}}_{r}} \|_{*} = \| \bm{Q}_{\widetilde{\bm{\mathcal{U}}}_{r}} (\bm{X}_r+\bm{H}) \bm{Q}_{\widetilde{\bm{\mathcal{V}}}_{r}} \|_{*}  \nonumber \\
&\ge \|   \bm{B}_L \bm{O}_L \bm{L} (\overline{\bm{X}}_{r} + \overline{\bm{H}}) \bm{R}^{\rm H} \bm{O}_R^{\rm H}  \bm{B}_R^{\rm H} \|_{*} - \|\bm{X}_{r}^{+} \|_{*}  \nonumber \\
&= \|  \bm{L} (\overline{\bm{X}}_{r} + \overline{\bm{H}}) \bm{R}^{\rm H} \|_{*} - \|\bm{X}_{r}^{+} \|_{*}  \nonumber \\
&= \Bigg\|
\begin{bmatrix}
\bm{L}_{11}\overline{\bm{X}}_{r,11}\bm{R}_{11}  & \\
&‌\bm{0}_{n-r}‌
\end{bmatrix} +‌\bm{L} \overline{\bm{H}} \bm{R}^{\rm H}
\Bigg\|_{*} - \|\bm{X}_{r}^{+} \|_{*}
 \end{align}
replacing these two upper and lower bounds in \eqref{A48} and considering convexity of nuclear norm we have:
\begin{align}
&\langle \bm{L} \overline{\bm{H}} \bm{R}^{\rm H}, \bm{G}^{\rm H} \rangle  \nonumber \\
&\le \Bigg\|
\begin{bmatrix}
\bm{L}_{11}\overline{\bm{X}}_{r,11}\bm{R}_{11}  & \\
&‌\bm{0}_{n-r}‌
\end{bmatrix} +‌\bm{L} \overline{\bm{H}} \bm{R}^{\rm H}
\Bigg\|_{*}  \nonumber \\
& - \Bigg\|
\begin{bmatrix}
\bm{L}_{11}\overline{\bm{X}}_{r,11}\bm{R}_{11}  & \\
&‌\bm{0}_{n-r}‌
\end{bmatrix}
\Bigg\|_{*} \le 0  \nonumber \\
 & \forall \bm{G} \in \partial\Bigg\|
 \begin{bmatrix}
 \bm{L}_{11}\overline{\bm{X}}_{r,11}\bm{R}_{11}  & \\
 &‌\bm{0}_{n-r}‌
 \end{bmatrix}
 \Bigg\|_{*} \label{A51}
\end{align}
In \eqref{A51}, $ \partial\| \bm A \|_*$ indicates the sub-differential of nuclear norm at point $\bm A$ \cite{recht2010guaranteed}. In order to fully describe sub-differential, let ${\rm rank}(\overline{\bm{X}}_{r,11})={\rm rank}({\bm X}_r) = r$ and for non-zero weights ${\rm rank}(\bm{L}_{11} \overline{\bm X}_{11} \bm{R}_{11})=r$. Consider svd for matrix $\bm{L}_{11} \overline{\bm X}_{r,11} \bm{R}_{11}$ as:
$$ \bm{L}_{11}\overline{\bm{X}}_{r,11}\bm{R}_{11} = \overline{\bm{U}}_r \overline{\bm{\Delta}}_r \overline{\bm{V}}_r^{\rm H} $$
Also let $\bm S$ be the sign matrix defined as:
\begin{align}
\bm{S}:= \begin{bmatrix}
\bm{S}_{11} & \\
& \bm{0}_{n-r}
\end{bmatrix} := \begin{bmatrix}
\overline{\bm{U}}_r \overline{\bm{V}}_r^{\rm H} & \\
& \bm{0}_{n-r}
\end{bmatrix} \label{S-matrix}
\end{align} 
Finally the sub-differential is defined as:
\begin{align}
&\partial\Bigg\|
\begin{bmatrix}
\bm{L}_{11}\overline{\bm{X}}_{r,11}\bm{R}_{11}  & \\
&‌\bm{0}_{n-r} ‌
\end{bmatrix}
\Bigg\|_{*}  \nonumber \\
&= \{\bm{G} \in \mathbb{R}^{n \times n } : \bm{G} = \begin{bmatrix}
	\bm{S}_{11} \in \mathbb{R}^{n \times n}& \\
	& \bm{G}_{22} \in \mathbb{R}^{(n-r) \times (n-r)}
	\end{bmatrix}  \nonumber \\
	&‌{\rm and } ~ \|\bm{G}_{22}\| \le 1\}  \nonumber \\
&\{ \bm{G} \in \mathbb{R}^{n \times n } : \mathcal{P}_{\overline{\bm{T}}}(\bm{G}) = \bm{S} =  \begin{bmatrix}
\bm{S}_{11} & \\
& \bm{0}_{n-r}
\end{bmatrix}   \nonumber \\
 & {\rm and }~  \mathcal{P}_{\overline{\bm{T}}^{\perp}}(\bm{G}) \le 1\} 
\end{align}
Considering \eqref{S-matrix} we have:
\begin{align}
&{\rm rank}(\bm{S}) = {\rm rank}(\bm{S}_{11}) =r, \|\bm{S}\| = \|\bm{S}_{11}\| = 1 \nonumber \\
&\|\bm{S}\|_{F} = \|\bm{S}_{11}\|_{F} = \sqrt{r}
\end{align}
Then according to all above equations and for $\|\bm{G}_{22} \| \leq 1$, \eqref{A51} will be changed to:
\begin{align}
\langle \bm{L}\overline{\bm{H}}\bm{R}^{\rm H} , \bm{S} +‌ \begin{bmatrix}
\bm{0}_{r} & \\
& \bm{G}_{22}
\end{bmatrix} \rangle \le 0 
\end{align}

\begin{align}
&0 \ge  \langle \bm{L}\overline{\bm{H}}\bm{R}^{\rm H} , \bm{S}  \rangle + \sup_{\|\bm{G}_{22}\| \le 1} \langle \bm{L}\overline{\bm{H}}\bm{R}^{\rm H} ,  \begin{bmatrix}
\bm{0}_{r} & \\
& \bm{G}_{22}
\end{bmatrix} \rangle  \nonumber \\
&= \langle \bm{L}\overline{\bm{H}}\bm{R}^{\rm H} , \bm{S}  \rangle + \sup_{\|\bm{G}\| \le 1} \langle \mathcal{P}_{\overline{\bm{T}}^{\perp}}(\bm{L}\overline{\bm{H}}\bm{R}^{\rm H}) ,  \bm{G} \rangle  \nonumber \\
&= \langle \bm{L}\overline{\bm{H}}\bm{R}^{\rm H} , \bm{S}  \rangle + \| \mathcal{P}_{\overline{\bm{T}}^{\perp}}(\bm{L}\overline{\bm{H}}\bm{R}^{\rm H}) \|_{*}   \nonumber \\
&= \langle \overline{\bm{H}} , \bm{L}^{\rm H}\bm{S}\bm{R}\rangle + \| \mathcal{P}_{\overline{\bm{T}}^{\perp}}(\bm{L}\overline{\bm{H}}\bm{R}^{\rm H}) \|_{*}  \nonumber \\
&=\Bigg\langle \overline{\bm{H}} , \begin{bmatrix}
\bm{L}_{11}\bm{S}_{11}\bm{R}_{11} & \bm{L}_{11}\bm{S}_{11}\bm{R}_{12} & \\
\bm{L}_{12}^{\rm H}\bm{S}_{11}\bm{R}_{11} &‌\bm{L}_{12}^{\rm H}\bm{S}_{11}\bm{R}_{12}& \\
& & \bm{0}_{n-2r}
\end{bmatrix}\Bigg\rangle  \nonumber \\
& +\Bigg\|
\begin{bmatrix}
\bm{0}_{r} &‌‌& & \\
&‌\bm{L}_{22}\overline{\bm{H}}_{22}\bm{R}_{22} & \bm{L}_{22}\overline{\bm{H}}_{23}\bm{\Gamma}_2 & \bm{L}_{22}\overline{\bm{H}}_{24} \\
&\bm{\Lambda}_2\overline{\bm{H}}_{32}\bm{R}_{22}& \bm{\Lambda}_2\overline{\bm{H}}_{33}\bm{\Gamma}_2 & \bm{\Lambda}_2\overline{\bm{H}}_{34}\\
&\overline{\bm{H}}_{42}\bm{R}_{22}& \overline{\bm{H}}_{34}\bm{\Gamma}_2 & \overline{\bm{H}}_{44}
\end{bmatrix}
\Bigg\|_{*}  \nonumber \\
& =\Bigg\langle \overline{\bm{H}} , \begin{bmatrix}
\bm{L}_{11}\bm{S}_{11}\bm{R}_{11} & \bm{L}_{11}\bm{S}_{11}\bm{R}_{12} & \\
\bm{L}_{12}^{\rm H}\bm{S}_{11}\bm{R}_{11} &‌\bm{0}_{r}& \\
& & \bm{0}_{n-2r}
\end{bmatrix}\Bigg\rangle  \nonumber \\
& + \langle \overline{\bm{H}}_{22}, \bm{L}_{12}^{\rm H}\bm{S}_{11}\bm{R}_{12} \rangle  \nonumber \\
& \Bigg\|
\begin{bmatrix}
\bm{0}_{r} &‌‌& & \\
&‌\bm{L}_{22}\overline{\bm{H}}_{22}\bm{R}_{22} & \bm{L}_{22}\overline{\bm{H}}_{23}\bm{\Gamma}_2 & \bm{L}_{22}\overline{\bm{H}}_{24} \\
 &\bm{\Lambda}_2\overline{\bm{H}}_{32}\bm{R}_{22}& \bm{\Lambda}_2\overline{\bm{H}}_{33}\bm{\Gamma}_2 & \bm{\Lambda}_2\overline{\bm{H}}_{34}\\
&\overline{\bm{H}}_{42}\bm{R}_{22}& \overline{\bm{H}}_{34}\bm{\Gamma}_2 & \overline{\bm{H}}_{44}
\end{bmatrix}
\Bigg\|_{*}  \nonumber \\
&=:\langle \overline{\bm{H}} , \overline{\bm{S}}^{\prime}\rangle +‌\langle \overline{\bm{H}}_{22}, \bm{L}_{12}^{\rm H}\bm{S}_{11}\bm{R}_{12} \rangle + \| \bm{L}^{\prime}\mathcal{P}_{\overline{\bm{T}}^{\perp}}(\overline{\bm{H}}) \bm{R}^{\prime} \|_{*} \label{A56}
\end{align} 
where \eqref{A56} is held due to the fact that spectral norm is the dual of nuclear norm. Also the matrices $\bm{S}'$ and $\bm{L}'$ are defined as below:
$$ \overline{\bm{S}}^{\prime} := \begin{bmatrix}
\bm{L}_{11}\bm{S}_{11}\bm{R}_{11} & \bm{L}_{11}\bm{S}_{11}\bm{R}_{12} & \\
\bm{L}_{12}^{\rm H}\bm{S}_{11}\bm{R}_{11} &‌\bm{0}_{r}& \\
& & \bm{0}_{n-r^{\prime}-r}
\end{bmatrix} $$
$$\bm{L}^{\prime} := \begin{bmatrix}
\bm{0}_{r} &  & \\
 &‌\bm{L}_{22}& & \\
 & & \bm{\Lambda}_2& \\
& & & \bm{I}_{n-r^{\prime}	-2r}
\end{bmatrix} $$
It is worth mentioning that $\overline{S}\in\overline{\bm T}$. Here are some properties of matrix $\overline{\bm S}'$:
\begin{align}
& \|‌ \overline{\bm{S}}^{\prime}  \|_{F} = \Bigg\|
\begin{bmatrix}
\bm{L}_{11}\bm{S}_{11}\bm{R}_{11} & \bm{L}_{11}\bm{S}_{11}\bm{R}_{12} & \\
\bm{L}_{12}^{\rm H}\bm{S}_{11}\bm{R}_{11} &‌\bm{0}_{r}& \\
& & \bm{0}_{n-2r}
\end{bmatrix}
\Bigg\|_{F}  \nonumber \\
& \le \Bigg\|
\begin{bmatrix}
\bm{L}_{11}\bm{S}_{11}\bm{R}_{11} & \bm{L}_{11}\bm{S}_{11}\bm{R}_{12} & \\
\bm{L}_{12}^{\rm H}\bm{S}_{11}\bm{R}_{11} & \bm{L}_{12}^{\rm H}\bm{S}_{11}\bm{R}_{12}& \\
& & \bm{0}_{n-2r}
\end{bmatrix}
\Bigg\|_{F} \nonumber \\
&‌\le \|[\bm{L}_{11} , \bm{L}_{12}]\| \|\bm{S}_{11}\|_{F} \|[\bm{R}_{11} , \bm{R}_{12}]\|  \nonumber \\
& = \|[\bm{L}_{11} , \bm{L}_{12}]\| \|\bm{S}\|_{F} \|[\bm{R}_{11} , \bm{R}_{12}]\|  \nonumber \\
& = \sqrt{r} \|[\bm{L}_{11} , \bm{L}_{12}]\| \|[\bm{R}_{11} , \bm{R}_{12}]\|  \nonumber \\
& =  \sqrt{r}  \sqrt{\max_i \big( \frac{\lambda_{1}(i)^4 \cos^2 \mathbf{\theta}_{u}(i) + \sin^2 \mathbf{\theta}_{u}(i)}{\lambda_{1}(i)^2 \cos^2 \mathbf{\theta}_{u}(i) + \sin^2 \mathbf{\theta}_{u}(i) }\big)}  \nonumber \\
& \quad \quad \quad \quad \quad \quad .\sqrt{\max_i \big( \frac{\gamma_{1}(i)^4 \cos^2 \mathbf{\theta}_{v}(i) + \sin^2 \mathbf{\theta}_{v}(i)}{\gamma_{1}(i)^2 \cos^2 \mathbf{\theta}_{v}(i) + \sin^2 \mathbf{\theta}_{v}(i) }\big)} \nonumber \\
&=: \sqrt{r} \alpha(\mathbf{\theta}_{u}(i),\mathbf{\theta}_{v}(i),\lambda_{1}(i),\gamma_{1}(i)) \label{A57}
\end{align}
The second and the last inequality of \eqref{A57} are held due to $\| \bm{AB}\|_F \leq \| \bm A \| \| \bm B \|_F$. Considering \eqref{A57}, \eqref{A56} will be written as:
\begin{align}
&0 \ge \langle \overline{\bm{H}} , \overline{\bm{S}}^{\prime} \rangle + \langle \overline{\bm{H}}_{22} , \bm{L}_{12}^{\rm H}\bm{S}_{11}\bm{R}_{12} \rangle + \| \bm{L}^{\prime}\mathcal{P}_{\overline{\bm{T}}^{\perp}}(\overline{\bm{H}}) \bm{R}^{\prime} \|_{*}  \nonumber \\
& \ge \langle \overline{\bm{H}} , \overline{\bm{S}}^{\prime} \rangle  - \|\mathcal{P}_{\overline{\bm{T}}^{\perp}}(\overline{\bm{H}}) \|_{*} \|\bm{L}_{12}\| \|\bm{S}_{11}\| \|\bm{R}_{12}\| + \|\mathcal{P}_{\overline{\bm{T}}^{\perp}}(\overline{\bm{H}}) \|_{*}  \nonumber \\
& - 
\| \mathcal{P}_{\overline{\bm{T}}^{\perp}}(\overline{\bm{H}}) -  \bm{L}^{\prime}\mathcal{P}_{\overline{\bm{T}}^{\perp}}(\overline{\bm{H}}) \bm{R}^{\prime} \|_{*}  \nonumber \\
& = \langle \overline{\bm{H}} , \overline{\bm{S}}^{\prime} \rangle + (1-\|\bm{L}_{12}\|\|\bm{R}_{12}\|)\|\mathcal{P}_{\overline{\bm{T}}^{\perp}}(\overline{\bm{H}}) \|_{*}  \nonumber \\
&- \|\mathcal{P}_{\overline{\bm{T}}^{\perp}}(\bm{I}_{n}) \mathcal{P}_{\overline{\bm{T}}^{\perp}}(\overline{\bm{H}})\mathcal{P}_{\overline{\bm{T}}^{\perp}}(\bm{I}_{n}) -  \bm{L}^{\prime}\mathcal{P}_{\overline{\bm{T}}^{\perp}}(\overline{\bm{H}}) \bm{R}^{\prime} \|_{*}  \nonumber \\
&‌\ge \langle \overline{\bm{H}} , \overline{\bm{S}}^{\prime} \rangle + (1-\|\bm{L}_{12}\|\|\bm{R}_{12}\|)\|\mathcal{P}_{\overline{\bm{T}}^{\perp}}(\overline{\bm{H}}) \|_{*}  \nonumber \\
& -  \|\mathcal{P}_{\overline{\bm{T}}^{\perp}}(\bm{I}_{n}) - \bm{L}^{\prime} \| \|\mathcal{P}_{\overline{\bm{T}}^{\perp}}(\overline{\bm{H}}) \|_{*}
\nonumber \\
& - \| \bm{L}^{\prime} \| \|\mathcal{P}_{\overline{\bm{T}}^{\perp}}(\overline{\bm{H}}) \|_{*} \|\mathcal{P}_{\overline{\bm{T}}^{\perp}}(\bm{I}_{n}) - \bm{R}^{\prime} \| \nonumber \\
& \ge \langle \overline{\bm{H}} , \overline{\bm{S}}^{\prime} \rangle + (1-\|\bm{L}_{12}\|\|\bm{R}_{12}\|)\|\mathcal{P}_{\overline{\bm{T}}^{\perp}}(\overline{\bm{H}}) \|_{*} \nonumber \\
& -  \Big\|\begin{bmatrix}
\bm{I}_{r} - \bm{L}_{22} & \\
& \bm{I}_{r} - \bm{\Lambda}_{2}
\end{bmatrix} \Big\| \|\mathcal{P}_{\overline{\bm{T}}^{\perp}}(\overline{\bm{H}}) \|_{*}
- \|\mathcal{P}_{\overline{\bm{T}}^{\perp}}(\overline{\bm{H}}) \|_{*} \nonumber \\
& \quad \quad \quad \quad \quad \quad \quad .\Big\|\begin{bmatrix}
\bm{I}_{r} - \bm{R}_{22} & \\
& \bm{I}_{r} - \bm{\Gamma}_{2}
\end{bmatrix} \Big\| \nonumber \\
& = \langle \overline{\bm{H}} , \overline{\bm{S}}^{\prime} \rangle  + (1-\|\bm{L}_{12}\|\|\bm{R}_{12}\|-  \Big\|\begin{bmatrix}
\bm{I}_{r} - \bm{L}_{22} & \\
& \bm{I}_{r} - \bm{\Lambda}_{2}
\end{bmatrix} \Big\| \nonumber \\
& -\Big\|\begin{bmatrix}
\bm{I}_{r} - \bm{R}_{22} & \\
& \bm{I}_{r} - \bm{\Gamma}_{2}
\end{bmatrix} \Big\|) \|\mathcal{P}_{\overline{\bm{T}}^{\perp}}(\overline{\bm{H}}) \|_{*} \nonumber \\
&  = \langle \overline{\bm{H}} , \overline{\bm{S}}^{\prime} \rangle + \Bigg(1- \sqrt{\max\Big(\frac{(1-\lambda_{1}(i)^{2})^2 \cos^2 \mathbf{\theta}_{u}(i) + \sin^2 \mathbf{\theta}_{u}(i)}{\lambda_{1}(i)^2\cos^2 \mathbf{\theta}_{u}(i) + \sin^2 \mathbf{\theta}_{u}(i)}\Big)} \nonumber \\
&\quad \quad \quad .\sqrt{\max\Big(\frac{(1-\gamma_{1}(i)^{2})^2 \cos^2 \mathbf{\theta}_{v}(i) \sin^2 \mathbf{\theta}_{v}(i)}{\gamma_{1}(i)^2\cos^2 \mathbf{\theta}_{v}(i) + \sin^2 \mathbf{\theta}_{v}(i)}\Big)} \nonumber \\
&-\max_i \{  \max_i (\lambda_{2}(i)-1) , \max_i \Big(\frac{\lambda_{1}(i)}{\sqrt{\lambda_{1}(i)^2\cos^2\mathbf{\theta}_{u}(i)+ \sin^2 \mathbf{\theta}_{u}(i)}}\nonumber\\
&-1 \Big)  \}  -\max_i \{  \max_i (\gamma_{2}(i)-1)  ,\nonumber\\
& \max_i \Big(\frac{\gamma_{1}(i)}{\sqrt{\gamma_{1}(i)^2\cos^2 \mathbf{\theta}_{v}(i) +\sin^2 \mathbf{\theta}_{v}(i)}}-1 \Big)  \}‌\Bigg
)\|\mathcal{P}_{\overline{\bm{T}}^{\perp}}(\overline{\bm{H}}) \|_{*}  \nonumber \\
& =: \langle \overline{\bm{H}} , \overline{\bm{S}}^{\prime} \rangle + (1-\alpha_6(\mathbf{\theta}_{u}(i),\mathbf{\theta}_{v}(i)\lambda_{i},\gamma_i)) ‌\|\mathcal{P}_{\overline{\bm{T}}^{\perp}}(\overline{\bm{H}}) \|_{*} \label{A58}
\end{align}
in which the first inequality holds due to \eqref{A56}, the second one comes from holder's inequality and the fact that $\overline{\bm H}_{22}$ is a part of $\mathcal{P}_{\overline{T}^{\perp}}(\overline{\bm H})$. The last inequality is due to $ab+a+b \leq \frac{3}{2}(a+b)$. 

Following lemma defines dual certificate.
\begin{lem} \label{lemmaA-2-6}
Let $\overline{\bm T}$ be the support of matrix $\overline{\bm X}_r$ as defined in \eqref{3-42},
let $\underset{i,j}{\rm min}~ p_{ij} \leq l$, consider $\overline{\mathcal{R} _\Omega}$ and $\overline{\mathcal{P}_p}$ as defined in \eqref{A40} and \eqref{A42}, respectively. As long as for $i,j \in [1:n]$ :
\begin{align}
 &\max [\log(\alpha_4n),1]. \frac{(\mu_i+\nu_j)r\log n}{n} \nonumber \\ 
&.\max[\alpha_5 \Big(1+ \max_{i} \frac{\breve{\mu}_i}{\mu_i} + \max_{j} \frac{\breve{\nu}_j }{\nu_j}\Big) ,1] \lesssim p_{ij} \le 1
\end{align}
then \eqref{A60} holds:
\begin{align} \|(\mathcal{P}_{\overline{\bm{T}}} - \mathcal{P}_{\overline{\bm{T}}}\circ \overline{\mathcal{A}}_p\circ\mathcal{P}_{\overline{\bm{T}}})(.) \|_{F \rightarrow F} \le \frac{1}{2} \label{A60}
\end{align}
and there exists a matrix $\overline{\Pi} \in \mathbb{R}^{n \times n}$ that ensures \eqref{A61} to \eqref{A63}
\begin{align}
&\| \overline{\bm{S}}^{\prime} - \mathcal{P}_{\overline{\bm{T}}}(\overline{\Pi}) \|_{F} \le \frac{l}{4\sqrt{2}} \label{A61}\\
&\mathcal{P}_{\overline{\bm{T}}^{\perp}}(\overline{\Pi}) \le \frac{1}{2} \label{A62}\\
&‌\overline{\Pi} = \overline{\mathcal{A}}_p(\overline{\Pi}) \label{A63}
\end{align}
where $\alpha_4$ and $\alpha_5$ can be defined as:
\begin{align}
&\alpha_4 = \alpha_4(u_i,v_i,\lambda_{1}(i),\gamma_{1}(i)):=  \nonumber \\
& \sqrt{\max_i \big( \frac{\lambda_{1}(i)^4 \cos^2 \mathbf{\theta}_{u}(i) + \sin^2 \mathbf{\theta}_{u}(i)}{\lambda_{1}(i)^2 \cos^2 \mathbf{\theta}_{u}(i) + \sin^2 \mathbf{\theta}_{u}(i) }\big)} \nonumber \\
&\sqrt{\max_i \big( \frac{\gamma_{1}(i)^4 \cos^2 \mathbf{\theta}_{v}(i) + \sin^2 \mathbf{\theta}_{v}(i)}{\gamma_{1}(i)^2 \cos^2 \mathbf{\theta}_{v}(i) + \sin^2 \mathbf{\theta}_{v}(i)}\big)} \nonumber \\
&\alpha_5 = \alpha_5(\mathbf{\theta}_{u}(i),\mathbf{\theta}_{v}(i),\lambda_{1}(i),\gamma_{1}(i)):= \nonumber \\
&\sqrt{\max_{i} (\lambda_{1}(i)^2\cos^2 \mathbf{\theta}_{u}(i) + \sin^2 \mathbf{\theta}_{u}(i))}.\nonumber \\
& \sqrt{\max_i \big( \frac{\gamma_{1}(i)^4 \cos^2 \mathbf{\theta}_{v}(i) + \sin^2 \mathbf{\theta}_{v}(i)}{\gamma_{1}(i)^2 \cos^2 \mathbf{\theta}_{v}(i) + \sin^2 \mathbf{\theta}_{v}(i) }\big)}.\Big( 1 + \sqrt{\tfrac{r^{\prime}+r}{r} \max_{i}\tfrac{\breve{\mu}_i}{\mu_i}}\Big) \nonumber \\
& +  \sqrt{\max_{i} (\gamma_{1}(i)^2\cos^2 \mathbf{\theta}_{v}(i) + \sin^2 \mathbf{\theta}_{v}(i)}\nonumber \\
&\sqrt{\max_i \big( \frac{\lambda_{1}(i)^4 \cos^2 \mathbf{\theta}_{u}(i) + \sin^2 \mathbf{\theta}_{u}(i)}{\lambda_{1}(i)^2 \cos^2 \mathbf{\theta}_{u}(i) + \sin^2 \mathbf{\theta}_{u}(i) }\big)}
.\Big( 1+ \sqrt{\tfrac{r^{\prime}+r}{r} \max_{i}\tfrac{\breve{\nu}_i}{\nu_i}}\Big) \nonumber \\
%& \alpha_6 = \alpha_4(\mathbf{\theta}_{u}(i),\mathbf{\theta}_{v}(i),\lambda_{1}(i),\gamma_{1}(i)) :=
\blacksquare
\end{align}
\end{lem}
So now, according to Lemma \ref{lemmaA-2-6} there exists a $\overline{\Pi}$ ensures \eqref{A61} to \eqref{A63}. Thus, \eqref{A56} can be re-written as:
\begin{align}
&0 \ge  \langle \overline{\bm{H}} , \overline{\bm{S}}^{\prime} \rangle + (1-\alpha_6) ‌\|\mathcal{P}_{\overline{\bm{T}}^{\perp}}(\overline{\bm{H}}) \|_{*} \nonumber \\
& = \langle \overline{\bm{H}} , \mathcal{P}_{\overline{\bm{T}}}(\overline{\Pi}) \rangle + \langle \overline{\bm{H}} , \overline{\bm{S}}^{\prime} -\mathcal{P}_{\overline{\bm{T}}}(\overline{\Pi}) \rangle + (1-\alpha_6) ‌\|\mathcal{P}_{\overline{\bm{T}}^{\perp}}(\overline{\bm{H}}) \|_{*} \nonumber \\ 
& = \langle \overline{\bm{H}} , \overline{\Pi} \rangle +  \langle \overline{\bm{H}} , \mathcal{P}_{\overline{\bm{T}}}(\overline{\Pi}) \rangle + \langle \overline{\bm{H}} , \overline{\bm{S}}^{\prime} -\mathcal{P}_{\overline{\bm{T}}}(\overline{\Pi}) \rangle 
 \nonumber \\
&‌+ (1-\alpha_6) ‌\|\mathcal{P}_{\overline{\bm{T}}^{\perp}}(\overline{\bm{H}}) \|_{*} \nonumber \\
& = -  \langle \overline{\bm{H}} , \mathcal{P}_{\overline{\bm{T}}^{\perp}}(\overline{\Pi}) \rangle + \langle \overline{\bm{H}} , \overline{\bm{S}}^{\prime} -\mathcal{P}_{\overline{\bm{T}}}(\overline{\Pi}) \rangle 
 \nonumber \\
&‌+ (1-\alpha_6) ‌\|\mathcal{P}_{\overline{\bm{T}}^{\perp}}(\overline{\bm{H}}) \|_{*} \nonumber \\
& \ge \|‌ \mathcal{P}_{\overline{\bm{T}}^{\perp}}(\overline{\bm{H}}) \|_{*} \| \mathcal{P}_{\overline{\bm{T}}^{\perp}}(\overline{\Pi}) \|-\|‌ \mathcal{P}_{\overline{\bm{T}}}(\overline{\bm{H}}) \|_{F} \| \overline{\bm{S}}^{\prime} -\mathcal{P}_{\overline{\bm{T}}}(\overline{\Pi})  \|_{F} \nonumber \\
&‌+ (1-\alpha_6) ‌\|\mathcal{P}_{\overline{\bm{T}}^{\perp}}(\overline{\bm{H}}) \|_{*} \nonumber \\
& \ge \frac{-1}{2} \|‌ \mathcal{P}_{\overline{\bm{T}}^{\perp}}(\overline{\bm{H}}) \|_{*} - \frac{l}{4\sqrt{2}} \|‌ \mathcal{P}_{\overline{\bm{T}}}(\overline{\bm{H}}) \|_{F} ‌+ (1-\alpha_6) ‌\|\mathcal{P}_{\overline{\bm{T}}^{\perp}}(\overline{\bm{H}}) \|_{*}  \nonumber \\
& = (\frac{1}{2} - \alpha_6) \|‌ \mathcal{P}_{\overline{\bm{T}}^{\perp}}(\overline{\bm{H}}) \|_{*} - \frac{l}{4\sqrt{2}} \|‌ \mathcal{P}_{\overline{\bm{T}}}(\overline{\bm{H}}) \|_{F} \label{A65}
\end{align}
for $\alpha_6 \leq 1$, \eqref{A65} is equivalent to:
\begin{align}
(\frac{1}{2} - \alpha_6) \|‌ \mathcal{P}_{\overline{\bm{T}}^{\perp}}(\overline{\bm{H}}) \|_{*} \le \frac{l}{4\sqrt{2}} \|‌ \mathcal{P}_{\overline{\bm{T}}}(\overline{\bm{H}}) \|_{F} \label{A66}
\end{align}
the third inequality of \eqref{A65} comes from Holder's inequality, also it's third inequality holds due to \eqref{A61} and \eqref{A62} and $\langle\overline{\bm H},\overline{\Pi}\rangle$
\begin{align}
\| \overline{\mathcal{A}}_p(\overline{\bm{H}}) \|_{F} = \| \mathcal{R} _\Omega(\bm{H}) \|_{F} = \| \mathcal{R} _\Omega(\widehat{\bm{X}} - \bm{X}) \|_{F} = 0 
\end{align}
\begin{align}
\langle \overline{\bm{H}} , \overline{\Pi} \rangle = \langle \overline{\bm{H}} , \overline{\mathcal{P}}_p(\overline{\Pi}) \rangle = \langle \overline{\mathcal{P}}_p(\overline{\bm{H}}) , \overline{\Pi} \rangle  =  0 
\end{align} 
In order to calculate upper bound of \eqref{A66} first consider:
\begin{align}
&\| \overline{\mathcal{A}}_p(\mathcal{P}_{\overline{\bm{T}}}(\overline{\bm{H}})) \|_{F} = \| \overline{\mathcal{A}}_p(\mathcal{P}_{\overline{\bm{T}}^{\perp}}(\overline{\bm{H}})) \|_{F}   \nonumber \\
&\le 
\| \overline{\mathcal{A}}_p(.) \|_{F \rightarrow F} \| \mathcal{P}_{\overline{\bm{T}}^{\perp}}({\bm{H}}) \| _{F}
\le \frac{1}{l} \| \mathcal{P}_{\overline{\bm{T}}^{\perp}}(\overline{\bm{H}}) \| _{F}\label{A69}
\end{align}
Now considering Lemma \ref{lemmaA-2-6}, one can say:
\begin{align}
&\| \overline{\mathcal{A}}_p(\mathcal{P}_{\overline{\bm{T}}}(\overline{\bm{H}})) \|_{F}^{2} = \langle \mathcal{P}_{\overline{\bm{T}}}(\overline{\bm{H}}) ,(\overline{\mathcal{A}}_p \circ \overline{\mathcal{A}}_p	)(\mathcal{P}_{\overline{\bm{T}}}(\overline{\bm{H}})) \rangle  \nonumber \\
&\ge \langle \mathcal{P}_{\overline{\bm{T}}}(\overline{\bm{H}}) , \overline{\mathcal{A}}_p	(\mathcal{P}_{\overline{\bm{T}}}(\overline{\bm{H}})) \rangle = \langle \mathcal{P}_{\overline{\bm{T}}}(\overline{\bm{H}}) , \mathcal{P}_{\overline{\bm{T}}}(\overline{\bm{H}}) \rangle  \nonumber \\
& + \langle \mathcal{P}_{\overline{\bm{T}}}(\overline{\bm{H}}) , ( \mathcal{P}_{\overline{\bm{T}}} \circ  \overline{\mathcal{A}}_p \circ \mathcal{P}_{\overline{\bm{T}}} - \mathcal{P}_{\overline{\bm{T}}}) \circ \mathcal{P}_{\overline{\bm{T}}}(\overline{\bm{H}}) \rangle \nonumber \\
& \ge \| \mathcal{P}_{\overline{\bm{T}}}(\overline{\bm{H}}) \|_{F}^{2} - \| \mathcal{P}_{\overline{\bm{T}}} \circ  \overline{\mathcal{A}}_p \circ \mathcal{P}_{\overline{\bm{T}}} - \mathcal{P}_{\overline{\bm{T}}} \|_{F \rightarrow F} \| \mathcal{P}_{\overline{\bm{T}}}(\overline{\bm{H}})\|_{F}^{2}
\nonumber \\
&\ge \frac{1}{2} \| \mathcal{P}_{\overline{\bm{T}}}(\overline{\bm{H}})\|_{F}^{2} \label{A70}
\end{align} 
Thus, comparing and combining \eqref{A69} and \eqref{A70} leads to:
\begin{align}
\| \mathcal{P}_{\overline{\bm{T}}}(\overline{\bm{H}})\|_{F} \le \frac{\sqrt{2}}{l} \| \mathcal{P}_{\overline{\bm{T}}^{\perp}}(\overline{\bm{H}})\|_{F} \label{A71}
\end{align}
Finally considering \eqref{A71}, \eqref{A66} leads to:
\begin{align}
(\frac{1}{2} - \alpha_6) \| \mathcal{P}_{\overline{\bm{T}}^{\perp}}(\overline{\bm{H}})\|_{*} \le \frac{l}{4\sqrt{2}} \| \mathcal{P}_{\overline{\bm{T}}^{\perp}}(\overline{\bm{H}})\|_{F} 
\end{align}
in which it can be concluded that as long as $\alpha_6 \leq \tfrac{1}{4}$, $\mathcal{P}_{{\overline{\bm T}}^\perp}(\overline{\bm H}) = 0$.
%$$  \mathcal{P}_{\overline{\bm{T}}^{\perp}}(\overline{\bm{H}}) = 0  $$ 
So According to the aforementioned tips, the error bound of $\overline{\bm H}$ is:
\begin{align}
&\| \overline{\bm{H}} \|_{F} \le  \| \mathcal{P}_{\overline{\bm{T}}^{\perp}}(\overline{\bm{H}}) \|_{F} + \| \mathcal{P}_{\overline{\bm{T}}}(\overline{\bm{H}}) \|_{F} \nonumber \\
& \le (\frac{\sqrt{2}}{l} +‌1) \| \mathcal{P}_{\overline{\bm{T}}^{\perp}}(\overline{\bm{H}}) \|_{F} = 0 ;
\end{align}
and $\overline{\bm X} = \bm X$. Expanding the existing results for low rank matrix and observations in noisy environments, one can say:
\begin{align}
&\| \widehat{\bm{X}} - \bm{X} \|_{F} \le \frac{\sqrt{h}}{l}   \| \bm{Q}_{\widetilde{\bm{\mathcal{U}}}_{r}} \bm{X}_{r^{+}} \bm{Q}_{\widetilde{\bm{\mathcal{V}}}_{r}} \|_{*} + \frac{e\sqrt{n}h^{\frac{3}{2}}}{l} \nonumber \\
&\frac{\sqrt{h}}{l} \| \bm{Q}_{\widetilde{\bm{\mathcal{U}}}_{r}}\| \| \bm{X}_{r^{+}} \|_{*} \| \bm{Q}_{\widetilde{\bm{\mathcal{V}}}_{r}}\| + \frac{e\sqrt{n}h^{\frac{3}{2}}}{l} \nonumber \\
& \frac{\sqrt{h}}{l} \| \bm{X}_{r^{+}} \|_{*} + \frac{e\sqrt{n}h^{\frac{3}{2}}}{l}
\end{align} 
\begin{align}
\| \widehat{\bm{X}} - \bm{X} \|_{F} \le  \frac{\sqrt{h}}{l} \| \bm{X}_{r^{+}} \|_{*} + \frac{e\sqrt{n}h^{\frac{3}{2}}}{l}
\end{align}
%\begin{align}
%&\max [\log(\alpha_4n),1]. \frac{(\mu_i+\nu_j)r\log n}{n} \nonumber \\ 
%&.\max[\alpha_5 \Big(1+ \max_{i} \frac{\breve{\mu}_i}{\mu_i} + \max_{j} \frac{\breve{\nu}_j }{\nu_j}\Big) ,1]	\preceq p_{ij} \le 1  \nonumber \\
%& \alpha_6 \le \frac{1}{4} 
%\end{align}
\end{proof}  
\section{Proof of Lemma \ref{lemmaA-2-6}}\label{proof lem 3}
As mentioned in the Lemma 3, $\bm{U}_r$ and $\widetilde{\bm{U}}_{r^{\prime}}$ are orthogonal bases of subspaces $\bm{\mathcal{U}}_{r} $ and $ \widetilde{\bm{\mathcal{U}}}_{r^{\prime}} $, respectively. Now without loss of  generality suppose that  
	\begin{align}
	\bm{U}_r^{\rm H} \widetilde{\bm{U}}_{r^{\prime}} = [\cos \mathbf{\bm{\theta}}_{u} \quad \bm{0}_{r \times r^{\prime}-r} ] :‌= \bm{U}_r^{\rm H}[\widetilde{\bm{U}}_{1,r} \quad \widetilde{\bm{U}}_{2,r^{\prime} - r}] \in \mathbb{R}^{r \times r^{\prime}} 
	\end{align}
where $ \widetilde{\bm{U}}_{1,r} \in \mathbb{R}^{n \times r} $ and  $ \widetilde{\bm{U}}_{2,r^{\prime} - r} \in \mathbb{R}^{ n \times r^{\prime} - r } $ are orthogonal bases for subspaces $ \widetilde{\bm{\mathcal{U}}}_{1,r} \subset \widetilde{\bm{\mathcal{U}}}_{r^{\prime}} $ and $ \widetilde{\bm{\mathcal{U}}}_{2,r^{\prime} -r } \subset \widetilde{\bm{\mathcal{U}}}_{r^{\prime}} $.

For construct orthonormal bases $\bm{B}_L$ in \eqref{B_l and B_r}we set  
	\begin{align}
	&\bm{U}^{\prime}_{1,r} := -(\bm{I} - \bm{U}_r\bm{U}_r^{\rm H}) \widetilde{\bm{U}}_{1,r}\sin ^{-1} (\mathbf{\bm{\theta}}_{u}) =  -\bm{P}_{\widetilde{\bm{\mathcal{U}}}^{\perp}_{r}}\widetilde{\bm{U}}_{1,r}\sin ^{-1} (\mathbf{\bm{\theta}}_{u}) \in \mathbb{R}^{n \times r}  \nonumber \\
 	&\bm{U}^{\prime}_{2,r^{\prime}-r} := -(\bm{I} - \bm{U}_r\bm{U}_r^{\rm H}) \widetilde{\bm{U}}_{2,r^{\prime}-r} = -\bm{P}_{\widetilde{\bm{\mathcal{U}}}^{\perp}_{r}}\widetilde{\bm{U}}_{2,r^{\prime}-r} \in \mathbb{R}^{n \times r^{\prime}-r}
% 	& \bm{U}^{\prime \prime} := -(\bm{I} - \bm{U}_r\bm{U}_r^{\rm H}) \widetilde{\bm{U}}^{\perp}_{2,r^{\prime}-r}\in \mathbb{R}^{n \times n-r^{\prime}-r}
 	\end{align}
 and consider  $$ {\rm span}(\bm{U}_{n-r^{\prime} - r}^{\prime \prime}) = {\rm span}([\bm{U}_r \quad \bm{U}_{r^{\prime}}^{\prime}])^{\perp}. $$
 Although we explained results and proof for column space but all results exist and honest for row spaces.

% 	\begin{align}
% 	\widetilde{\bm{U}}_r^{\prime} := (\bm{I} - \widetilde{\bm{U}}_r\widetilde{\bm{U}}_r^{\rm H}) \bm{U}_r(\sin u)^{-1} \in \mathbb{R}^{r \times r}
% 	\end{align} 

%%%%%Span property 	
% 	\begin{align}
% 	&{\rm span}([\bm{U}_r \quad \widetilde{\bm{U}}_r])  \nonumber \\
% 	&={\rm span}([\bm{U}_r \quad (\bm{I} - \bm{U}_r\bm{U}_r^{\rm H})\widetilde{\bm{U}}_r])  \nonumber \\
% 	& = {\rm span}([\bm{U}_r \quad -(\bm{I} - \bm{U}_r\bm{U}_r^{\rm H})\widetilde{\bm{U}}_r\begin{bmatrix}
% 	(\sin u)^{-1} & \\
% 	& \bm{I}_{r^{\prime}-r}
% 	\end{bmatrix}])  \nonumber \\
% 	&= {\rm span}([\bm{U}_r \quad \bm{U}_{r^{\prime}}^{\prime}])
% 	\end{align}
% 	\begin{align*}
% 	&{\rm span}([\bm{U}_r \quad \bm{U}_{r^{\prime}}^{\prime}]) = {\rm span}([\widetilde{\bm{U}}_r \quad \widetilde{\bm{U}}_{r^{\prime}}^{\prime}]) \nonumber \\
% 	&={\rm span}([\bm{U}_r \quad \widetilde{\bm{U}}_{r^{\prime}}])
% 	\end{align*}
\section{Proof of Lemma \ref{lem 4}}\label{proof lemma 4}
We use from tow important following point for proof of equalities in Lemma \ref{lem 4}:
\begin{enumerate}
	\item The operator norm of the diagonal matrix is the max element of the matrix. 
	\item For $ \bm{X} \in \mathbb{R}^{n \times n }$
	$$ \| \bm{X} \| =  \sqrt{\lambda_{\max}(\bm{X}^{\rm H}\bm{X})} = \sigma_{\max}(\bm{X}),$$  
	where $\lambda_{\max}(\bm{X}^{\rm H}\bm{X})$ is largest eigenvalue of $\bm{X}^{\rm H}\bm{X}$ and $\sigma_{\max}(\bm{X})$ the largest
	singular value of $\bm{X}$ \cite[Lemma A.5.]{foucart2017mathematical}
\end{enumerate}

%Note that the operator norm of the diagonal matrix is the max element of the matrix. So the operator norm of $ \bm{L}_{11}$, $\bm{L}_{12}$ and $\bm{I}_r - \bm{L}_{22}$ is equal to max elements of them.
	\begin{align}
	& \| \bm{L}_{11} \| = \| \bm{\Delta}_L\| = \sqrt{\max_{i} (\lambda_{1}(i)^2\cos^2 \mathbf{\theta}_{u}(i) + \sin^2 \mathbf{\theta}_{u}(i)},  \nonumber \\
	& \|\bm{L}_{12}\| = \sqrt{\max_{i}\Big(\frac{(1-\lambda_{1}(i)^{2})^2 \cos^2 \mathbf{\theta}_{u}(i) + \sin^2 \mathbf{\theta}_{u}(i)}{\lambda_{1}(i)^2\cos^2 \mathbf{\theta}_{u}(i) + \sin^2 \mathbf{\theta}_{u}(i)}\Big)},   \nonumber \\
	& \|\bm{I}_r - \bm{L}_{22} \| = \| \bm{I}_r - \bm{\Lambda}\bm{\Delta}_L^{-1}\| = \nonumber \\
	& \sqrt{\max_i\Big(\frac{(\lambda_{1}(i)-\sqrt{\lambda_i^2\cos^2 \mathbf{\theta}_{u}(i) + \sin^2 \mathbf{\theta}_{u}(i)})^2}{\lambda_{1}(i)^2\cos^2 \mathbf{\theta}_{u}(i) + \sin^2 \mathbf{\theta}_{u}(i)}\Big)},  \nonumber \\
		& \Big\|\begin{bmatrix}
	 \bm{I}_{r} - \bm{L}_{22} & \\
	& \bm{I}_{r} - \bm{\Lambda}_{2}
	\end{bmatrix} \Big\| =  \max_i \{  \max_i (1- \lambda_{2}(i))\nonumber \\
	&‌ \quad \quad \quad \quad \quad \quad , \max_i \Big(1- \frac{\lambda_{1}(i)}{\sqrt{\lambda_{1}(i)^2\cos^2 \mathbf{\theta}_{u}(i) + \sin^2 \mathbf{\theta}_{u}(i)}} \Big) \}, \nonumber \\
	&\|[\bm{L}_{11} \quad  \bm{L}_{12}]\|^2 = \max_i \Bigg\|\Bigg[
	 \sqrt{(\lambda_{1}(i)^2\cos^2 \mathbf{\theta}_{u}(i) + \sin^2 \mathbf{\theta}_{u}(i)} \nonumber \\
	 &\quad \quad  \quad \quad \quad \quad  \quad \quad \frac{(1-\lambda_{1}(i)^2) \cos \mathbf{\theta}_{u}(i) \sin \mathbf{\theta}_{u}(i)}{\sqrt{(\lambda_{1}(i)^2\cos^2 \mathbf{\theta}_{u}(i) + \sin^2 \mathbf{\theta}_{u}(i)}}
	\Bigg]\Bigg\|^2_{2} \nonumber \\
	&= \max_i \big( \frac{\lambda_{1}(i)^4 \cos^2 \mathbf{\theta}_{u}(i) + \sin^2 \mathbf{\theta}_{u}(i)}{\lambda_{1}(i)^2 \cos^2 \mathbf{\theta}_{u}(i) + \sin^2 \mathbf{\theta}_{u}(i) }\big)  \nonumber \\
	&\Bigg\|\begin{bmatrix}
	\bm{0}_r & \bm{L}_{12} & &  \\
	& \bm{L}_{22}-\bm{I}_{r} & & \\
	& & \bm{\Lambda}_2 - \bm{I}_{r^{\prime}-r} & \\
	& & &\bm{0}_{n-r^{\prime}-r}
	\end{bmatrix}\Bigg\|^{2} = \nonumber \\
	&\max \Big\{ \Bigg\| \begin{bmatrix}
	\bm{L}_{12} \\  \bm{L}_{22}-\bm{I}_{r}
	\end{bmatrix}\Bigg\|_2^2  , \| \bm{\Lambda}_2 - \bm{I}_{r^{\prime}-r} \|_2^2 \Big\}	
	\nonumber \\
	& =\max_i \{[ \max_i \Big(1- (\frac{\lambda_{1}(i)}{\sqrt{\lambda_{1}(i)^2\cos^2 u_i + \sin^2 u_i}})^2 + \nonumber \\
	&\quad \quad (\frac{(1-\lambda_{1}(i))^2\cos^2 \mathbf{\theta}_{u}(i)\sin^2 \mathbf{\theta}_{u}(i)}{\lambda_{1}(i)^2\cos^2 \mathbf{\theta}_{u}(i) + \sin^2 \mathbf{\theta}_{u}(i)}) \Big) , \max_i (\lambda_{2}(i)-1)^2] \}.
	\end{align}
	\end{appendices}
%	\ifCLASSOPTIONcaptionsoff
%	\newpage
%	\fi
%	\bibliographystyle{ieeetr}
%	\bibliography{MyrefrenceMRadMC}
	\bibliographystyle{ieeetr}
	\bibliography{MyrefrenceMRadMC}

\begin{thebibliography}{10}

\bibitem{candes2009exact}
E.~J. Cand{\`e}s and B.~Recht, ``Exact matrix completion via convex
  optimization,'' {\em Foundations of Computational mathematics}, vol.~9,
  no.~6, p.~717, 2009.

\bibitem{recht2011simpler}
B.~Recht, ``A simpler approach to matrix completion.,'' {\em Journal of Machine
  Learning Research}, vol.~12, no.~12, 2011.

\bibitem{candes2010matrix}
E.~J. Candes and Y.~Plan, ``Matrix completion with noise,'' {\em Proceedings of
  the IEEE}, vol.~98, no.~6, pp.~925--936, 2010.

\bibitem{recht2010guaranteed}
B.~Recht, M.~Fazel, and P.~A. Parrilo, ``Guaranteed minimum-rank solutions of
  linear matrix equations via nuclear norm minimization,'' {\em SIAM review},
  vol.~52, no.~3, pp.~471--501, 2010.

\bibitem{gross2010quantum}
D.~Gross, Y.-K. Liu, S.~T. Flammia, S.~Becker, and J.~Eisert, ``Quantum state
  tomography via compressed sensing,'' {\em Physical review letters}, vol.~105,
  no.~15, p.~150401, 2010.

\bibitem{haldar2010spatiotemporal}
J.~P. Haldar and Z.-P. Liang, ``Spatiotemporal imaging with partially separable
  functions: A matrix recovery approach,'' in {\em 2010 IEEE International
  Symposium on Biomedical Imaging: From Nano to Macro}, pp.~716--719, IEEE,
  2010.

\bibitem{zhao2010low}
B.~Zhao, J.~P. Haldar, C.~Brinegar, and Z.-P. Liang, ``Low rank matrix recovery
  for real-time cardiac mri,'' in {\em 2010 IEEE International Symposium on
  Biomedical Imaging: From Nano to Macro}, pp.~996--999, IEEE, 2010.

\bibitem{srebro2010collaborative}
N.~Srebro and R.~R. Salakhutdinov, ``Collaborative filtering in a non-uniform
  world: Learning with the weighted trace norm,'' in {\em Advances in Neural
  Information Processing Systems}, pp.~2056--2064, 2010.

\bibitem{aravkin2014fast}
A.~Aravkin, R.~Kumar, H.~Mansour, B.~Recht, and F.~J. Herrmann, ``Fast methods
  for denoising matrix completion formulations, with applications to robust
  seismic data interpolation,'' {\em SIAM Journal on Scientific Computing},
  vol.~36, no.~5, pp.~S237--S266, 2014.

\bibitem{bennett2007netflix}
J.~Bennett, S.~Lanning, {\em et~al.}, ``The netflix prize,'' in {\em
  Proceedings of KDD cup and workshop}, vol.~2007, p.~35, New York, 2007.

\bibitem{so2007theory}
A.~M.-C. So and Y.~Ye, ``Theory of semidefinite programming for sensor network
  localization,'' {\em Mathematical Programming}, vol.~109, no.~2-3,
  pp.~367--384, 2007.

\bibitem{daei2018optimal}
S.~Daei, A.~Amini, and F.~Haddadi, ``Optimal weighted low-rank matrix recovery
  with subspace prior information,'' {\em arXiv preprint arXiv:1809.10356},
  2018.

\bibitem{eftekhari2018weighted}
A.~Eftekhari, D.~Yang, and M.~B. Wakin, ``Weighted matrix completion and
  recovery with prior subspace information,'' {\em IEEE Transactions on
  Information Theory}, vol.~64, no.~6, pp.~4044--4071, 2018.

\bibitem{ardakani2020multi}
H.~S.~F. Ardakani, S.~Daei, and F.~Haddadi, ``Multi-weight nuclear norm
  minimization for low-rank matrix recovery in presence of subspace prior
  information,'' {\em arXiv preprint arXiv:2005.10878}, 2020.

\bibitem{rao2015collaborative}
N.~Rao, H.-F. Yu, P.~Ravikumar, and I.~S. Dhillon, ``Collaborative filtering
  with graph information: Consistency and scalable methods.,'' in {\em NIPS},
  vol.~2, p.~7, Citeseer, 2015.

\bibitem{angst2011generalized}
R.~Angst, C.~Zach, and M.~Pollefeys, ``The generalized trace-norm and its
  application to structure-from-motion problems,'' in {\em 2011 International
  Conference on Computer Vision}, pp.~2502--2509, IEEE, 2011.

\bibitem{jain2013provable}
P.~Jain and I.~S. Dhillon, ``Provable inductive matrix completion,'' {\em arXiv
  preprint arXiv:1306.0626}, 2013.

\bibitem{xu2013speedup}
M.~Xu, R.~Jin, and Z.-H. Zhou, ``Speedup matrix completion with side
  information: Application to multi-label learning,'' in {\em Advances in
  neural information processing systems}, pp.~2301--2309, 2013.

\bibitem{mohan2010reweighted}
K.~Mohan and M.~Fazel, ``Reweighted nuclear norm minimization with application
  to system identification,'' in {\em Proceedings of the 2010 American Control
  Conference}, pp.~2953--2959, IEEE, 2010.

\bibitem{zhou2012kernelized}
T.~Zhou, H.~Shan, A.~Banerjee, and G.~Sapiro, ``Kernelized probabilistic matrix
  factorization: Exploiting graphs and side information,'' in {\em Proceedings
  of the 2012 SIAM international Conference on Data mining}, pp.~403--414,
  SIAM, 2012.

\bibitem{fathi2021two}
H.~Fathi, E.~Rangriz, and V.~Pourahmadi, ``Two novel algorithms for low-rank
  matrix completion problem,'' {\em IEEE Signal Processing Letters}, vol.~28,
  pp.~892--896, 2021.

\bibitem{ardakani2019greedy}
H.~Ardakani, S.~Fazael, S.~Daei, and F.~Haddadi, ``A greedy algorithm for
  matrix recovery with subspace prior information,'' {\em arXiv preprint
  arXiv:1907.11868}, 2019.

\bibitem{donoho2006compressed}
D.~L. Donoho, ``Compressed sensing,'' {\em IEEE Transactions on information
  theory}, vol.~52, no.~4, pp.~1289--1306, 2006.

\bibitem{candes2008restricted}
E.~J. Candes, ``The restricted isometry property and its implications for
  compressed sensing,'' {\em Comptes rendus mathematique}, vol.~346, no.~9-10,
  pp.~589--592, 2008.

\bibitem{needell2017weighted}
D.~Needell, R.~Saab, and T.~Woolf, ``Weighted-minimization for sparse recovery
  under arbitrary prior information,'' {\em Information and Inference: A
  Journal of the IMA}, vol.~6, no.~3, pp.~284--309, 2017.

\bibitem{chen2015completing}
Y.~Chen, S.~Bhojanapalli, S.~Sanghavi, and R.~Ward, ``Completing any low-rank
  matrix, provably,'' {\em The Journal of Machine Learning Research}, vol.~16,
  no.~1, pp.~2999--3034, 2015.

\bibitem{foucart2017mathematical}
S.~Foucart and H.~Rauhut, ``A mathematical introduction to compressive
  sensing,'' {\em Bull. Am. Math}, vol.~54, pp.~151--165, 2017.

\end{thebibliography}
\end{document}